\documentclass{aa}  

\usepackage[normalem]{ulem}  
\usepackage{graphicx}  
\usepackage{txfonts}   
\usepackage{xcolor}    
\usepackage{orcidlink}
\usepackage{multirow}
\usepackage{hyperref}  
\hypersetup{
    colorlinks=true,
    urlcolor=blue,
    linkcolor=blue,
    citecolor=blue
}
\def \MSUN{\rm M_{\odot}}  
\newcommand{\mytilde}{\raise.19ex\hbox{$\scriptstyle\sim$}} 
\begin{document}

\title{Star-Forming vs. Quenched Galaxies in Voids: Insights into the Role of Mergers }
\titlerunning{Starforming VS Quenched Galaxies in Void}  

\author{
    Mohammad Reza Shojaei\, \orcidlink{0009-0004-7055-2203} \inst{1},
    Saeed Tavasoli\, \orcidlink{0000-0003-0126-8554}\inst{1},
    Parsa Ghafour\, \orcidlink{0009-0003-2960-1563}\inst{1}
}
\authorrunning{Mohammad Reza Shojaei et al.}  

\institute{
    \inst{1} Department of Astronomy and High Energy Physics, Kharazmi University, No.43 South Mofateh St, Tehran, Iran \\ 
    \email{\href{mohammadreshojaei@gmail.com}{mohammadreshojaei@gmail.com}}
}

\date{Received XX; accepted XX}

 
  \abstract
   {Cosmic voids, the largest under-dense structures in the Universe, are crucial for exploring galaxy evolution. These vast, sparsely populated regions are home to void galaxies—predominantly gas-rich, star-forming, and blue—that evolve more slowly than those in denser environments. Additionally, the correlation between galaxy mergers and specific properties of galaxies, such as the star formation rate, is not fully understood, particularly in these under-dense environments. }
   {We aim to analyze the evolutionary histories of star-forming and quenched void galaxies across cosmic time (\( z \leq 2 \)). By investigating their stellar mass assembly, we aim to calculate the formation time of these void galaxies. We will also statistically examine the rates of major, minor, and all mergers (any stellar mass ratio) at different evolutionary timescales. Additionally, we will explore how these evolutionary paths and merger rates influence the star formation rates of void galaxies. 

 }
   {We select galaxies from the highest-resolution version of cosmological magneto-hydrodynamical simulations
of galaxy formation in IllustrisTNG, specifically from the \(302.6^3\) Mpc\(^3\) box size, focusing on those with stellar masses \(\geq 10^8 M_{\odot}\). These galaxies are classified into star-forming and quenched categories using UVJ diagrams and specific star formation rate (sSFR) thresholds. We also select void galaxies using the 'AM' algorithm. To analyze merger rates, we utilize public catalogs that contain comprehensive information and statistics on the merging history of all galaxies throughout the timeline of the TNG Illustris simulations. }
   {Quenched void galaxies exhibit high star formation rates (SFR) in high redshifts, significantly decreasing at lower redshifts (\( z \lesssim 0.5 \)). These galaxies have higher dark matter halos than star-forming galaxies across all redshifts, leading to rapid gas consumption. They formed earlier and experienced more major mergers in earlier epochs but fewer recent mergers, resulting in a lack of fresh gas for sustained star formation. Also, star-forming and high-mass quenched void galaxies show higher SFRs in mergers compared to non-merger galaxies.
}
{The study highlights that formation time, merger rates, and dark matter halos play a crucial role in the star formation history of void galaxies. Rapid and earlier time gas consumption due to earlier formation time and the absence of recent mergers could lead to quenched void galaxies at lower redshifts, providing valuable insights into galaxy evolution in low-density environments. }
\keywords{{galaxies: evolution – galaxies: interactions – galaxies: star formation}}

\maketitle
\section{Introduction}
Understanding the effects of galaxy mergers is essential for developing a comprehensive view of hierarchical galaxy evolution within the Lambda cold dark matter (\(\Lambda\)CDM) framework(\cite{kauffmann1993formation}; \cite{navarro1996structure};  \cite{hopkins2010mergers}; \cite{conselice2014evolution}; \cite{somerville2015physical}).  Apart from secular evolution through accreting gas from filaments, galaxies can significantly increase their stellar mass (\(M_\star\)) and halo mass by merging with other galaxies (\cite{blumenthal1984formation}; \cite{davis1985evolution}; \cite{neistein2006natural}; \cite{cattaneo2011galaxies}; \cite{rodriguez2015merger}), as well as significantly alter their structure through morphological transformation (\cite{toomre1972galactic};\cite{mihos1995gasdynamics}; \cite{lotz2010effect};\cite{hambleton2011advanced}). Furthermore, cosmological hydrodynamic galaxy simulations suggest that more stellar material comes from mergers and orbiting satellites (\cite{rodriguez2016stellar}; \cite{pillepich2018simulating}).

A variety of mechanisms have been proposed to enhance our understanding of the physical origins behind star formation quenching (see e.g. the review by \cite{somerville2015physical}), including (i) the heating of the inner halo gas through cosmological accretion due to ram pressure drag and local shocks (i.e. gravitational quenching; \cite{dekel2008gravitational}), (ii) the stability of discs against fragmentation to bound clumps (i.e. morphological quenching; \cite{martig2009morphological}; \cite{gensior2020heart}; \cite{gensior2020heart}), (iii) the removal of the gas supply resulting from active galactic nuclei (AGN) activity, and/or stellar feedback (e.g. \cite{springel2005modelling}; \cite{bower2006breaking}; \cite{sijacki2007unified}; \cite{cattaneo2009role}; \cite{fabian2012observational}), (iv) the interaction between the galaxy gas with the intracluster medium in high-density environments (i.e. environmental or satellite quenching; \cite{gunn1972infall};  \cite{larson1980evolution};  \cite{moore1998morphological};  \cite{bekki2009ram};  \citep{peng2010mass,peng2012mass}, and (v) the interaction with other galaxies (i.e. major mergers; \cite{di2005energy}; \cite{croton2006many}; \cite{hopkins2008cosmological}; \cite{somerville2008semi}).

Observations support the picture that merger-induced instabilities drive gas flows into the center of merger remnants (\cite{blumenthal2018go}), dilute the gas-phase metallicity (\cite{rupke2010galaxy}; \cite{torrey2012metallicity}; \cite{bustamante2018merger}), and enhance the star formation rate (SFR) on either nuclear or global scales (\cite{ellison2008galaxy}; \cite{robotham2014galaxy}; \cite{sparre2022gas}). Numerical simulations of galaxy pairs provide an ideal framework to clarify the merger process and its effects on the galaxies involved, confirming that these interactions and mergers increase the star formation rate (SFR) (\cite{di2007star}; \cite{cox2008effect}; \cite{scudder2012galaxy}; \cite{perret2014evolution}; \citep{moreno2015mapping, moreno2021spatially}) and regulate the gas content (\cite{mihos1995gasdynamics}; \citet{sinha2009numerical}; \cite{sparre2022gas}) in the nuclear region (\cite{moreno2021spatially}). Research indicates that the typical increase in star-formation rate (SFR) of a merger is, at most, a factor of two, much lower than what would typically be considered a starburst (\cite{ellison2013galaxy}; \cite{knapen2015interacting}; \cite{silva2018galaxy}). Mergers can potentially enhance the activity of an active galactic nucleus (e.g., \cite{sanders1996luminous}; \cite{ellison2019definitive}). However, more recent studies indicate that this may not necessarily be true in every instance (e.g. \cite{darg2010galaxy}; \cite{weigel2018fraction}). 

Gas-rich (wet) mergers can sustain greater star formation rates because there is ample fuel for producing new stars (\cite{lin2008redshift}; \cite{perez2011chemical}; \cite{athanassoula2016forming}). However, gas-poor (dry) mergers do not have gas readily available, so it is harder to form starbursts in these systems (\cite{bell2006dry}; \cite{naab2006properties}; \citet{lin2008redshift}). In dense environments, dry galaxy mergers are more common than wet mergers because they are more gas-poor than gas-rich galaxies (\citet{lin2010wet}). The percentage of dry mergers also increases as the age of the Universe increases (\cite{lin2008redshift}). Gas-poor galaxies dominate at high masses (stellar mass \(\gtrsim 10^{10.7} \MSUN\)), and as a result, mergers involving two high-mass galaxies usually result in dry interactions, which can suppress star formation (\citet{robotham2014galaxy}).

Numerous prior studies have sought to identify the key factors that influence star formation variability in mergers, including the stellar mass ratio of the two galaxies, their initial orbital parameters, projected separation, total stellar mass, bulge to total mass ratio, and environment (\cite{di2007star}; \cite{cox2008effect}; \cite{ellison2008galaxy}; \cite{scudder2012galaxy}; \cite{torrey2012metallicity}; \cite{perret2014evolution}; \cite{domingue2016major}; \cite{pan2018effect}; \citet{jia2021relations}). The mass ratio of stellar mergers is a crucial parameter used to differentiate between major and minor mergers, traditionally set at 0.25 (\citet{robotham2013galaxy}; \citet{pan2018effect}). Designating the more massive member as the primary galaxy and the less massive one as the secondary, \cite{davies2015galaxy} found that primary galaxies enhance their star formation rate, regardless of whether they experience major or minor mergers. 

The study by \cite{sol2006effects} , utilizing the 2-degree Field Galaxy Redshift Survey (2dFGRS) and the Sloan Digital Sky Survey (SDSS), demonstrated that galaxy interactions are particularly effective at triggering significant star formation activity in low- and moderate-density environments. They found that the enhancement of star formation in major galaxy pairs is notably higher in low-density environments. Therefore, examining how galaxy mergers influence the star formation rates of galaxies in low-density environments is essential. 

Understanding the importance of mergers in the assembly history of galaxies requires studying galaxy merger rates as a function of cosmic time (\cite{carlberg2000caltech}; \cite{patton2002dynamically}; \cite{conselice2003evidence}; \citep{lin2004deep2,lin2008redshift}; \citet{lotz2008evolution}; \cite{de2009vimos}; \citet{bluck2009surprisingly}) and understanding the level of triggered star formation during galaxy interactions (\cite{lambas2003galaxy}; \cite{nikolic2004star}; \cite{woods2007minor}; \cite{lin2007aegis}; \cite{barton2007isolating}).

One of the main challenges in merger studies is the difficulty of identifying a large sample of merging galaxies. Visually detecting galaxies is time-consuming and often yields inconsistent results. Different observers may classify the same galaxy in various ways, and even the same observer might assign different labels on other days. Different mergers have been simulated, allowing us to study the SFRs of the merging galaxies throughout the entire merger sequence from the first passage to coalescence (e.g., \cite{springel2005modelling}; \cite{hopkins2006unified}; \cite{randall2008constraints}; \cite{rupke2010galaxy}). These simulations have shown that SFR is enhanced when the merging galaxies are close to one another at first pass, second pass, and coalescence (\cite{moreno2019interacting}). 

In recent years, it has become possible to make large-volume, cosmological and hydrodynamical simulations of galaxy formation and evolution with statistically significant and increasingly more realistic galaxy populations, using, for example, the EAGLE (Evolution and Assembly of Galaxies and their Environments) simulation suite (\cite{crain2015eagle}; \cite{schaye2015eagle}), the Illustris simulations (\citep{vogelsberger2014introducing,vogelsberger2014properties}; \cite{genel2014introducing}), Simba (\cite{dave2019simba}), or IllustrisTNG (\cite{springel2018first}; \cite{pillepich2018simulating}). These simulations, which employ a comprehensive galaxy formation model and state-of-the-art numerical code, are successful in reproducing a wide range of observations, including the cosmic star-formation history, stellar population properties, stellar mass functions, scaling relations, clustering properties, galaxy sizes, and morphologies (\cite{furlong2015evolution}; \cite{rodriguez2015merger};\cite{sparre2017star}; \citet{pillepich2018simulating};\cite{nelson2019illustristng}).

The population of void galaxies is typically characterized by low-mass, blue, star-forming galaxies with young stellar populations (\citep{rojas2004photometric, rojas2005spectroscopic}, \citep{hoyle2005luminosity,hoyle2012photometric}; \cite{tavasoli2015galaxy}; \cite{moorman2016star}; \citet{florez2021void}; \citet{jian2022star}; \citet{rodriguez2024evolutionary}). One possible explanation is that galaxies in these environments may experience weaker gravitational forces due to a lower density of matter. This decrease in gravity could result in a slower rate of star formation, leading to smaller stellar masses (\cite{ricciardelli2014star};  \citet{dominguez2023galaxies}). 

Another potential outcome of living in void environments is that these galaxies are likely to be more isolated, possess fewer nearby galaxies, and have limited gas available for star formation. Thus, these galaxies might exhibit unique morphologies and chemical compositions compared to denser areas. Consequently, by examining the characteristics of galaxies in cosmic voids, we can gain enhanced insights into the formation and evolution of galaxies, along with the universe's large-scale structure. Void galaxies remain unaffected by transformation processes (like ram-pressure stripping, starvation, and harassment) that occur in group and cluster environments (\cite{beygu2017void}; \citet{argudo2024morphologies}; \cite{rodriguez2024evolutionary}). 

In this study, we utilize the IllustrisTNG simulation to examine the evolutionary pathways and merger rate histories of star-forming and quenched galaxies in void environments.  While the merging of galaxies in voids has received limited attention in previous research, this paper presents the first comprehensive statistical analysis of merger rates- distinguishing between minor and major mergers- across various evolutionary stages of both star-forming and quenched galaxies in voids.

IllustrisTNG provides a series of cosmological, gravity + magnetohydrodynamics simulations of galaxies in representative portions of synthetic universes carried out in a periodic box of 302.6 Mpc on a side (\cite{springel2018first}; \citet{marinacci2018first}; \citet{nelson2018first}; \citet{naiman2018first}; \citet{pillepich2018simulating}). Because of the large volume covered by the simulation, the self-consistent treatment of baryons, and the physically motivated galaxy formation model used (\cite{vogelsberger2013model}), the IllustrisTNG Simulation provides a unique opportunity to study the galaxy-galaxy merger rate with unprecedented precision and physical fidelity. We also use the “Merger History” catalog (\citet{rodriguez2017role} and \citet{eisert2023ergo}), which contains information and statistics about the merging history of all subhalos (i.e., galaxies) across time.

This paper is organized as follows. Section 2 overviews the IllustrisTNG simulations and describes the methodology used to identify galaxies within voids, classify them as quenched or star-forming, and apply specific star formation rates and UVJ diagram thresholds. In Section 3, we present the results of our analysis, including the evolutionary behavior of galaxy properties, the mass assembly history of galaxies, and merger statistics. Section 4 offers a detailed discussion of our findings and concludes with a summary of the results.

\section{Methods}
\subsection{The IllustrisTNG simulations}
The IllustrisTNG simulations (\citet{naiman2018first}, \citet{marinacci2018first}, \citet{springel2018first}, \cite{pillepich2018simulating}, \cite{nelson2018first}), referred to as TNG, are a series of advanced cosmological magnetohydrodynamical simulations designed to investigate key physical processes vital for galaxy formation and evolution. These simulations build and improve upon the original Illustris project (\cite{vogelsberger2013model}, \cite{torrey2015synthetic}, \cite{vogelsberger2014introducing}, \cite{genel2014introducing}, \cite{sijacki2015illustris}, \cite{nelson2015illustris}), modifying and adding numerous features to improve the agreement between the simulation and observational results. They provide a comprehensive physical model of galaxy formation from the early Universe to the present (\cite{weinberger2018supermassive}, \cite{pillepich2018first}).

The simulation suite contains three volumes: TNG50, TNG100, and TNG300, with sizes of \(51.7^3\), \(110.7^3\), and \(302.6^3\) comoving Mpc\(^3\), respectively. The TNG simulations adopt the cosmological parameters from Planck 2016, assuming \(\Omega_{\Lambda,0} = 0.6911\), \(\Omega_{m,0} = 0.3089\), \(\Omega_{b,0} = 0.0486\), \(\sigma_8 = 0.8159\), \(n_s = 0.9667\), and \(h = 0.6774\). 

This work focuses on results from the highest-resolution version of TNG300, in which the dark matter (DM) and stellar particle/mean baryonic gas cell mass are \(5.9 \times 10^7\) and \(1.1 \times 10^7 \, \MSUN\), respectively. TNG300 has \(2 \times 2500^3\) initial resolution elements with a gravitational softening length of 1.48 kpc at \(z=0\). This represents an approximately one-order-of-magnitude decrease in mass resolution and an increase in the gravitational softening length by a factor of two compared to TNG100.
The simulations are performed with the AREPO code (\cite{springel2010moving}), which solves Poisson’s equation for gravity by employing a tree-particle-mesh algorithm. The code uses the finite volume method for magnetohydrodynamics in the simulation domain's unstructured, moving Voronoi tessellation. Poisson’s gravity equation is solved by employing a Tree-Particle Mesh (TREE-PM; \cite{xu1994new}; \cite{bode2000tree}; \cite{bagla2002treepm}) that computes the contribution of short—and long-range forces using its tree and particle-mesh algorithms, respectively. Voronoi gas cells are particles at their center of mass, along with all other matter components.

\begin{figure}[t] 
    \centering
    \includegraphics[width=0.5\textwidth]{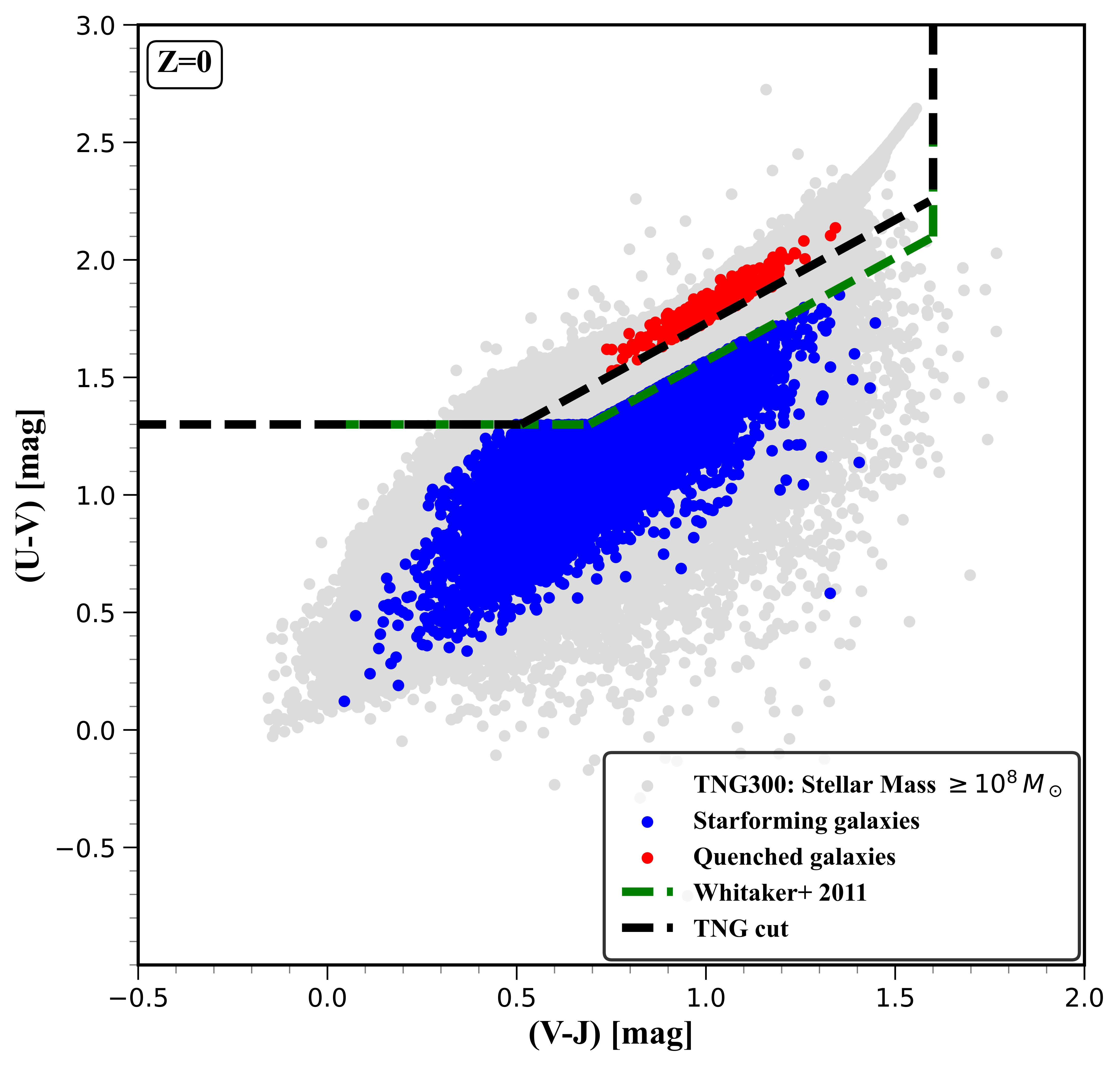} 
    \caption{The UVJ diagram at redshift \(z=0\) shows the distinction between star-forming (blue dots) and quenched galaxies (red dots). The green dashed line represents the classification threshold from Whitaker et al. (2011), while the black dashed line follows (Donnari2019). Grey points represent the total number of void galaxies with a stellar half-mass radius \(\geq 10^8 \, \MSUN\).}
    \label{fig:UVJ_diagram} 
\end{figure}

TNG uses the SUBFIND algorithm (\cite{davis1985evolution}, \cite{springel2001populating}, \cite{dolag2009substructures}) to identify gravitationally bound structures within the simulation. These structures consist of subhaloes and dark matter haloes, with baryonic components that make up galaxies within each subhalo. Subhaloes are tracked using SUBLINK merger trees (\cite{rodriguez2015merger}). A subhalo's descendant is the one that shares the highest weighted sum of individual particles—gas, stars, and dark matter—with its progenitor. These particles are ranked by gravitational binding energy and weighted by $(\text{rank})^{-1}$. A merger occurs when multiple subhaloes have a shared descendant, with the main progenitor identified as the subhalo with the most massive history (\cite{de2006formation}).

\subsection{Merger History in the TNG}
In this research, we use a catalog of merger histories (\cite{rodriguez2017role}, and \cite{eisert2023ergo}) that includes data and statistics regarding the merger history of all subhalos (i.e., galaxies) throughout all time. Here, mergers are split into three categories: major (stellar mass ratio > 1/4), minor (stellar mass ratio between 1/10 and 1/4), and all (any stellar mass ratio) mergers. The mass ratio is always determined by the stellar masses of the two merging galaxies when the secondary galaxy reaches its maximum stellar mass. To avoid spurious flyby and re-merger events, mergers are only included when both galaxies can be traced back to a time when each belonged to a different FoF group. A cleaning method has also been implemented to filter out mergers involving events attributed to non-cosmological subhalos ("clumps"). Initially, any subhalos marked with SubhaloFlag == False are excluded. Furthermore, the secondary is disregarded if absent for over two snapshots.

\subsection{Galaxy sample}
\subsubsection{Finding galaxies}
This paper uses the publicly available data from the most oversized simulation box (\cite{nelson2019illustristng}) with the best resolution at this level, TNG300-1 (hereafter TNG300), in the $z = 0$ snapshot. The properties of haloes and subhaloes utilized in this study are derived from the application of the Friends-of-Friends (FoF) and SUBFIND algorithms (\cite{davis1985evolution}, \cite{springel2001populating}), which are employed to detect substructures and, consequently, galaxies within the simulated volumes. In this study, we obtain the stellar masses by summing the masses of all stellar particles located within twice the stellar half-mass radius ($M_\star \)) as outlined by \cite{pillepich2018first}. We subsequently refined our analysis to identify all subhaloes that satisfy the established criterion for stellar mass $M_\star \geq 10^8 \, M_\odot$. With these prescriptions, we gather approximately 587,000 galaxies in TNG300 at $z = 0$.

\subsubsection{Void Identification}
Cosmic voids constitute the most under-dense regions in the universe. Identifying these regions is challenging because their definitions rely on factors such as shape, density thresholds, and the tracers used. The literature offers various ways to define them ([e.g.,\cite{kirshner1981million}, \cite{kauffmann1991voids}, \cite{sahni1994evolution}, \cite{benson2003galaxy}, \cite{padilla2005spatial}, \cite{platen2007cosmic}, \cite{neyrinck2008zobov}, \cite{lavaux2010precision}, \cite{sutter2015vide}). Overall, they all exhibit similar characteristics for galaxies located within these voids. The populations of void galaxies are generally composed of low-mass, blue, star-forming galaxies with young stellar populations (\cite{rojas2004photometric}, \cite{hoyle2005luminosity}, \cite{rojas2005spectroscopic}, \cite{hoyle2012photometric}, \cite{tavasoli2015galaxy}, \cite{moorman2016star}, \cite{florez2021void}, \cite{jian2022star}).

We aimed to study galaxies' evolution histories and statistical merger rates in void environments. For identified void galaxies from the initial data, we employed the void finder algorithm introduced by \cite{aikio1998simple}  (subsequently referred to as the AM algorithm), which has been enhanced to a three-dimensional version (\cite{tavasoli2013challenge}). We aim to analyze the merger history of central galaxies; thus, we distinguish between central and satellite galaxies. Our final count of central void galaxies is about 48,500 in different stellar masses.
The distinction between quenched and star-forming galaxies is a significant area of research. It is often ambiguous and somewhat arbitrary from theoretical and observational perspectives. Therefore, in this study, we aim to make a substantial contribution by simultaneously exploring two different classification criteria: the UVJ Diagram and the Specific Star Formation Rate threshold, to differentiate these two populations of galaxies more effectively.

\begin{figure*}[t] 
    \centering
    \includegraphics[width=\textwidth, keepaspectratio]{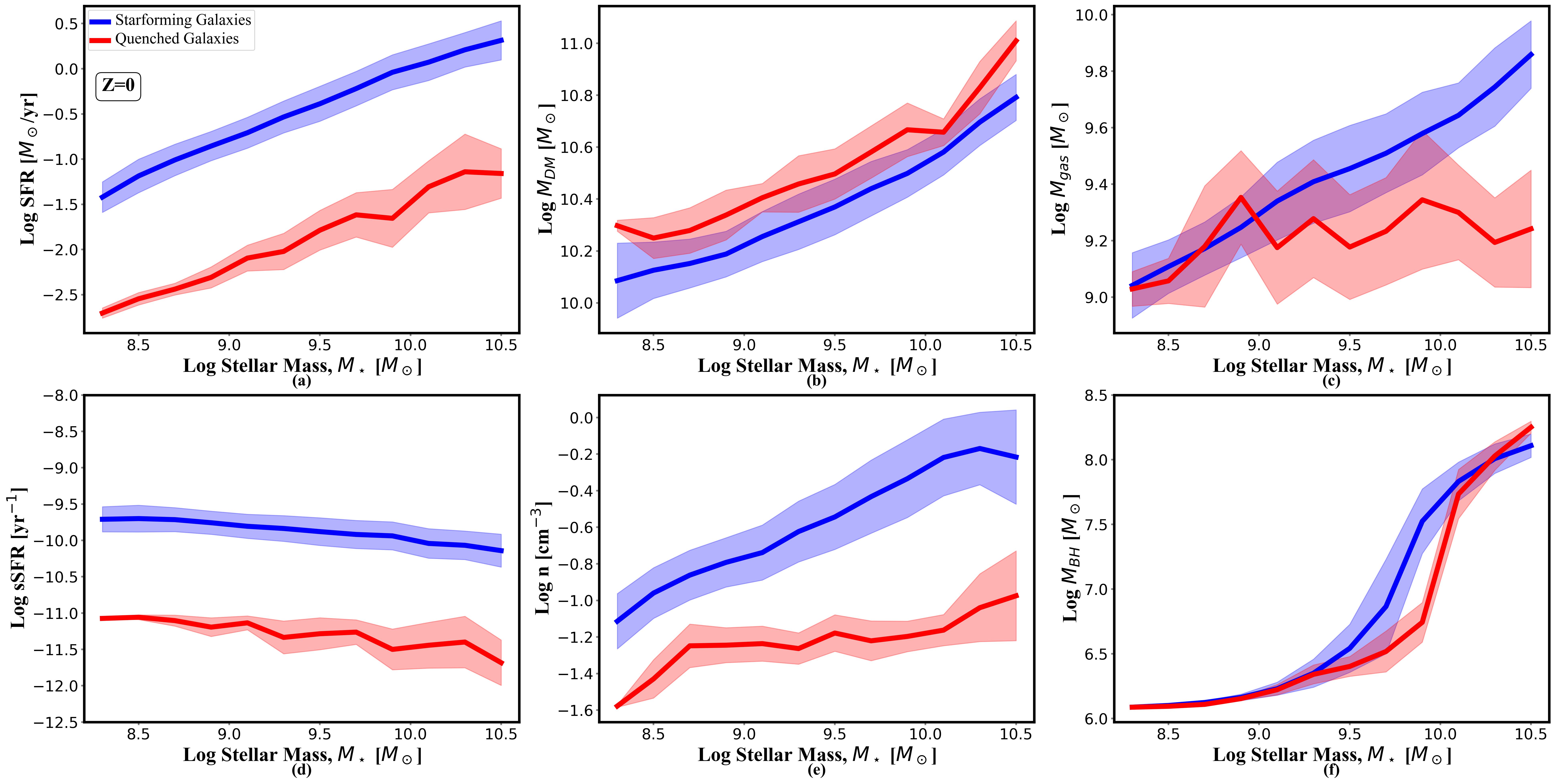}
    \caption{Comparative analysis of six key parameters: star formation rate(SFR), the mass of dark matter (\(M_\mathrm{DM}\)), the mass of gas (\(M_\mathrm{gas}\),) specific star formation rate (sSFR), the average number density of gas (\(n\)) within twice the half-mass stellar radius and supermassive black hole mass( \(M_\mathrm{BH}\)) for star-forming (blue lines) and quenched (red lines) galaxies as a function of stellar mass in void environments at \(z=0\). Lines indicate the median value, and shaded regions represent each parameter's Median Absolute Error (MAE).}
    \label{fig:stellar_mass_2comparison}
\end{figure*}

\subsubsection{Galaxy Classification Using the UVJ Diagram}
A significant method for identifying quiescent galaxies commonly employed in observations is the color-color diagram, especially the rest-frame U-V versus V-J plane. (Hereafter, UVJ diagram;\cite{wuyts2011star};\citep{williams2009detection,williams2010evolving}; \cite{whitaker2010age}; \cite{patel2013hst}; \cite{quadri2011tracing}). For plotting the UVJ diagram, we use the “SDSS ugriz and UVJ Photometry/Colors with Dust” catalog in TNG300 at redshift \(z=0\). This catalog contains synthetic stellar photometry (i.e., colors), including the effects of dust obscuration, corresponding to the fiducial dust model of \cite{nelson2018first} (i.e., Model C). Specifically, we utilize the rest-frame UVJ catalog as analyzed by \cite{donnari2019star}. 

 We adopt two different thresholds to separate star-forming and quenched galaxies based on UVJ cuts; first, apply the criteria derived from \cite{whitaker2011newfirm}:
\[
(U - V ) > 1.3 \quad \& \quad (V - J) < 1.6, \quad \text{valid for} \; 0 < z < 1.5.
\]
Secondly, we utilize the best estimate from \cite{donnari2019star} to differentiate TNG galaxies in the UVJ diagram:
\[
(U - V ) > 0.88 \times (V - J) + 0.85, \quad \text{applied for} \; 0 \leq z \leq 1.
\]
Fig.1 shows star-forming and quenched void galaxies in blue and red dots, respectively. This figure's green and black dashed lines represent the thresholds derived from  \cite{whitaker2011newfirm} and \cite{donnari2019star}.  We also eliminated the overlap region (the "small sample") between these two criteria to ensure a more confident and precise separation. This adjustment minimizes ambiguity and enhances the reliability of our results by focusing on galaxies that unambiguously fall into either category.
\subsubsection{Specific Star Formation Rate Threshold}
In the TNG model, star formation follows the approach described by \cite{springel2003cosmological}. Gas is converted into star particles in a stochastic manner when its density exceeds \(n_H = 0.1\, \mathrm{cm^{-3}}\). This conversion occurs over a timescale empirically determined to align with the Kennicutt-Schmidt relation (\cite{kennicutt1989star}). For each galaxy's star-formation rate (SFR), we use the sum of the instantaneous SFRs of all the gas cells assigned twice the stellar half-mass radius. In the literature, we commonly define "quenched" galaxies as those whose logarithmic Specific Star Formation Rate (sSFR) falls below a certain fixed threshold at any redshift, namely \(\mathrm{sSFR} \leq 10^{-11}\, \mathrm{yr^{-1}}\). Star-forming galaxies are thus those with sSFR larger than this threshold. Such threshold-based separation is often used in observations (\cite{mcgee2011dawn}, \cite{wetzel2013galaxy}, \cite{lin2014environment}, \cite{jian2018first}). After applying the UVJ and sSFR thresholds, the figure 1 illustrates the final sample selection of our star-forming and quenched void galaxies, represented in blue and red, respectively. Moreover, for our galaxies with \(M_\star \geq 10^8 \, \MSUN\), we also do not consider any values for \(\mathrm{sSFR} \leq 10^{-12.5}\, \mathrm{yr^{-1}}\)(\cite{donnari2019star})
Finally these thresholds precisely identify 9,137 star-forming and 249 quenched void galaxies in the TNG300 simulation at z=0, enabling a focused analysis of their evolutionary pathways and merger rate histories.

\section{Results}
\subsection{Star-forming and quenched galaxies as a function of Stellar Mass}

Before examining the merger histories of star-forming and quenched galaxies in void environments, we compare the essential characteristics of star-forming and quenched galaxies in voids at \(z=0\) using the TNG300 Illustris simulation. In Figure 2, we present a comparative analysis of the properties of star-forming and quenched void galaxies with stellar mass($M_\star \)) ranging from \(10^{8.2} \, M_\odot\) to \(10^{10.6} \, M_\odot\). We examine the logarithms of six parameters: star formation rate (SFR), dark matter mass (\(M_\mathrm{DM}\)), gas mass (\(M_\mathrm{gas}\)),  specific star formation rate (\(\mathrm{sSFR} \equiv \mathrm{SFR}/M_\mathrm{stars}\)), gas number density (\(n\)) within twice the half-mass stellar radius of each galaxy, and the mass of the supermassive black hole (SMBH) (\(M_\mathrm{BH}\)), as a function of stellar mass. The blue and red solid lines indicate the median relation of each parameter versus stellar mass for star-forming and quenched void galaxies. To quantify the dispersion, we also utilize shaded error bars to represent the Median Absolute Error (MAE) for each parameter in each mass bin of our samples.

\begin{figure*}[t] 
    \centering
    \includegraphics[width=\textwidth, keepaspectratio]{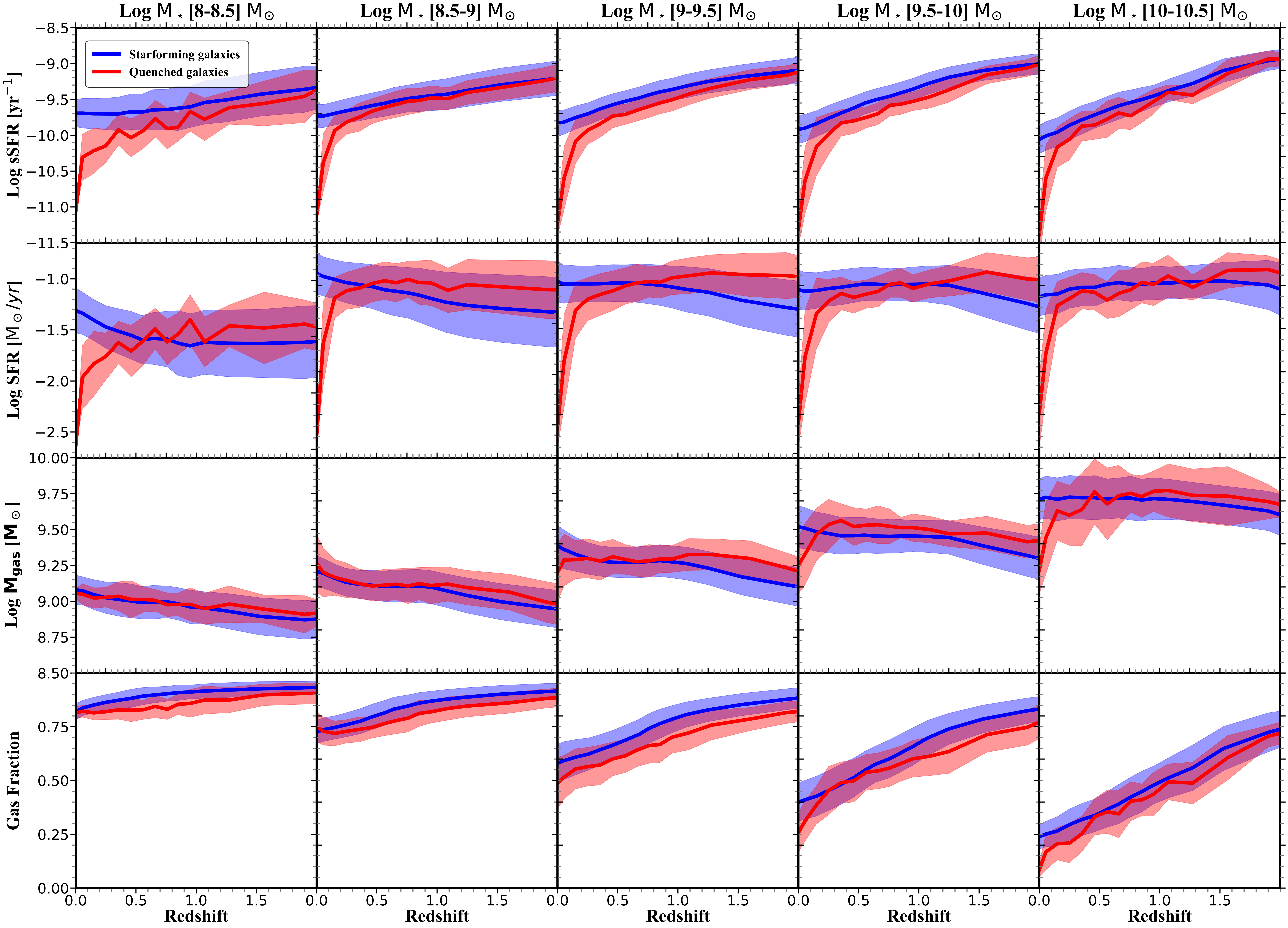}
   \caption{Evolution of specific star formation rate (sSFR), star formation rate (SFR), gas mass ($M_\text{gas}$), and gas fraction as a function of redshift ($z \leq 2$) for star-forming (red) and quenched (blue) galaxies in the IllustrisTNG 300 simulation. Lines in each panel represent the median of each parameter within redshift bins, while the shaded regions indicate the absolute error in the median, reflecting uncertainties in the measurements }

    \label{fig:stellar_mass_3comparison}
\end{figure*}
In Figure 2(a), we present the relation between the star formation rate (SFR) and stellar mass ($M_\star \)) for each central galaxy in both the star-forming and quenched samples at \(z=0\). Galaxies are defined based on their ‘instantaneous’ SFR, the sum of the SFRs of all gas cells that are gravitationally bound and within twice the stellar half-mass radius. As shown in previous studies (\citep{donnari2019star,donnari2021quenched}), the SFR increases with stellar mass, and star-forming galaxies exhibit higher SFR than quenched galaxies across all stellar mass bins.
\begin{figure*}[t] 
    \centering
    \includegraphics[width=\textwidth, keepaspectratio]{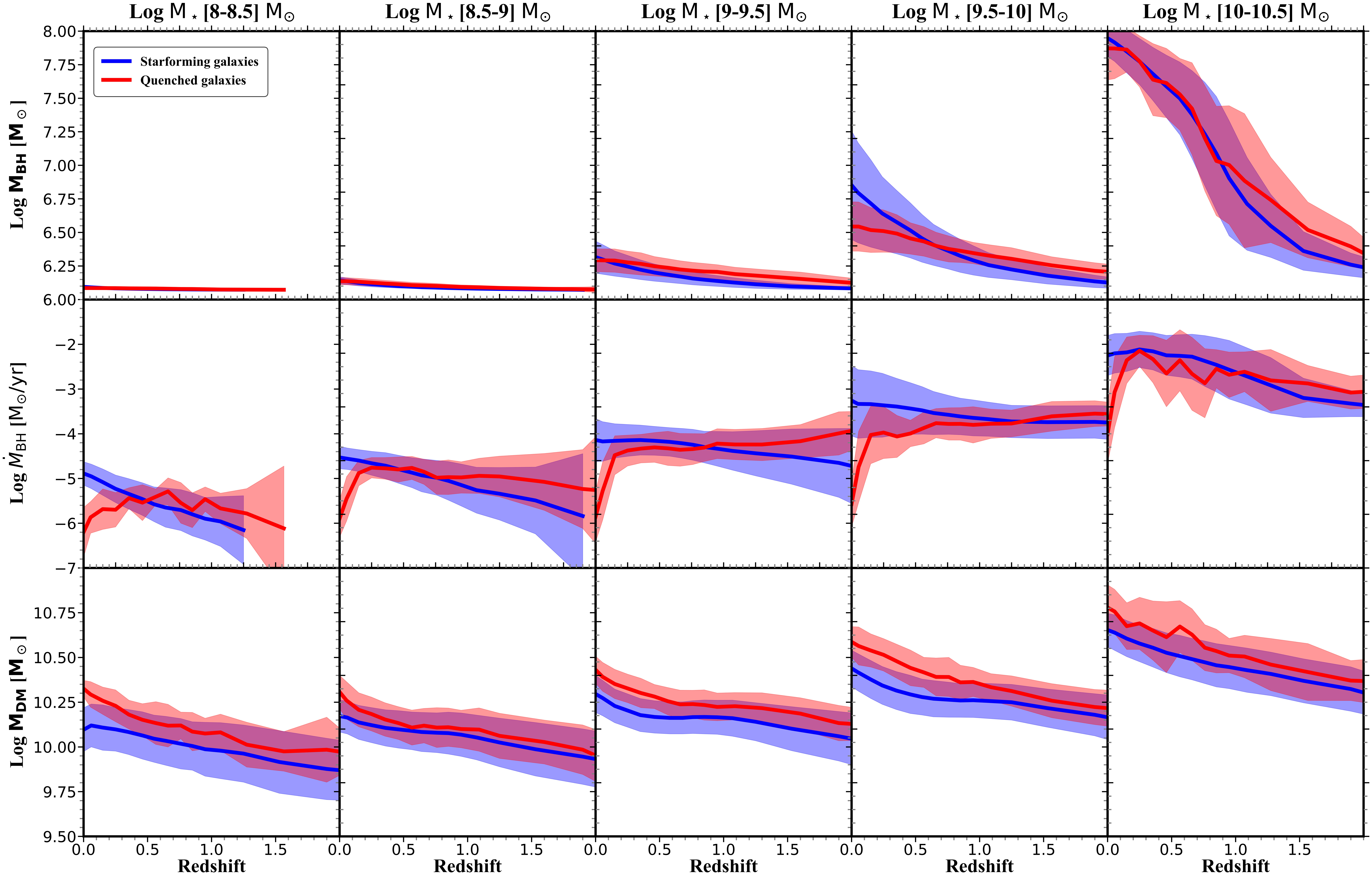}
   \caption{Evolution of black hole mass ($\log M_\text{BH}$), black hole accretion rate ($\log \dot{M}_\text{BH}$), and dark matter halo mass ($\log M_\text{DM}$) as a function of redshift ($z \leq 2$) for star-forming (blue) and quenched (red) galaxies in the IllustrisTNG 300 simulation. Lines in each panel represent the median of each parameter within redshift bins, while the shaded regions indicate the absolute error in the median, reflecting uncertainties in the measurements  }

    \label{fig:stellar_mass_4comparison}
\end{figure*}

Figure 2(b) shows the relation between dark matter mass (\(M_{\rm DM}\)) and stellar mass (\(M_{\rm stars}\)) for both star-forming and quenched void galaxies. Our findings indicate that quenched galaxies have slightly higher dark matter mass than star-forming galaxies of the same stellar mass. This suggests quenched galaxies may have formed in more massive dark matter halos.
Figure 2(c) shows the mass of gas in these galaxies that is calculated as the sum of gas particles in twice the stellar mass radius of galaxies and indicates that at low-mass galaxies \([10^{8.2}-10^{9}] \, M_{\odot}\) have slightly the same, and in higher stellar mass \([10^{9}-10^{10.6}] \, M_{\odot}\), star-forming galaxies have higher gas mass than quenched galaxies. As the stellar mass increases to \([10^{9}-10^{10.6}] \, M_{\odot}\), we observe a separation between the median lines, where the star-forming galaxies (blue line) consistently have higher gas masses than the quenched galaxies (red line). 

In Figure 2(d), Star-forming galaxies consistently have higher specific star formation rates (sSFR) than quenched galaxies across all mass bins, demonstrating our thresholds for classifying star-forming and quenched void galaxies. In TNG simulations, the multi-phase interstellar medium (ISM) star formation and pressurization are treated following the \cite{springel2003cosmological} model. Specifically, gas above a star formation density threshold of \(n > 0.1 \, \mathrm{cm}^{-3}\) forms stars stochastically following the empirically defined Kennicutt-Schmidt relation and assuming a Chabrier (\cite{chabrier2003galactic}) initial mass function. The gas cell number density is calculated as the average of all gas cells located within the half-mass radius of each galaxy, as shown in Figure 2(e). Quenched galaxies exhibit a lower gas number density than star-forming galaxies across all mass categories, and this discrepancy increases in higher stellar mass.   

Supermassive black hole (SMBH) feedback operates in two primary modes: thermal (or quasar) mode and kinetic (or radio) mode. Thermal feedback typically occurs when the black hole is in a high-accretion state, releasing energy by heating the surrounding gas. In contrast, kinetic feedback generally occurs when the black hole is in a low-accretion state, driving powerful jets that displace gas from the galaxy center (\cite{weinberger2016simulating}, \cite{pillepich2018simulating}, \cite{li2020correlations}). According to \cite{weinberger2018supermassive}, black holes with masses up to about \(10^{8.5} \, M_\odot\) are most effective in the thermal mode, where energy is primarily released as heat, preventing gas from cooling and condensing into stars. This gradual heating process suppresses star formation.   

Figure 2(f) shows the black hole mass–stellar mass relation for our quenched and star-forming samples which agrees well with observational data from \cite{savorgnan2016supermassive}.  
Black holes accrete at low rates at low stellar masses, leading to slower growth than stellar mass growth. As the stellar mass increases, sufficient gas accumulates around the black hole, resulting in higher accretion rates and eventually allowing the black hole to grow more rapidly. This accelerates black hole growth, consistent with the accretion rate–black hole mass relation, $\dot{M}_{\rm BH} \propto M_{\rm BH}^2$ \cite{weinberger2016simulating}. All SMBHs in our sample have \(M_{BH} \leq 10^{8.2} \, M_\odot\); therefore, they are in thermal feedback mode. In IllustrisTNG simulation, a black hole accreting in the thermal mode injects pure thermal energy into the surrounding gas particles. \cite{weinberger2018supermassive} shows that the total amount of thermal energy released in this mode can be large yet has little effect on cooling and the galaxy’s SFR. Thermally injected energy, as implemented for black hole thermal mode feedback in TNG, is more likely to be immediately radiated away rather than have any lasting impact on the thermodynamic properties of the gas, especially for dense gas where the cooling times are short (\cite{navarro1993simulations}; \cite{katz1995cosmological}). These low-mass SMBHs are always dominated by growth via accretion in the thermal mode, with mergers being a second, sub-dominant growth channel. Also, star-forming void galaxies have higher black hole mass in the mass range  \([10^{9.5}-10^{10}] \, M_{\odot}\).  One possible explanation is that an SMBH grows through galaxy mergers (\cite{volonteri2005rapid}), which supply the SMBH with a new gas reservoir and may star-forming void galaxies experience higher merger rate leading to increased gas accretion for black hole growth. 

\subsection{Evolutionary Analysis of Galaxy Properties}
Understanding the evolutionary history of galaxies is crucial for unraveling the processes that govern their formation over cosmic time. The evolution of key galaxy properties, such as star formation rate (SFR), specific star formation rate (sSFR), gas mass ($M_\text{gas}$), the gas fraction \(gas   fraction \equiv M_{\text{gas}}/(M_{\text{gas}} + M_{\text{star}}) \), the mass of dark matter ($M_\text{DM}$) within twice the stellar half-mass radius, the mass of supermassive black hole($M_\text{BH}$) and black hole accretion rate($\dot{M}_\text{BH}$), provides insights into the underlying physical mechanisms that drive galaxy growth and quenching. These properties are fundamental when studied in void environments, where the interplay between the local environment and galaxy evolution can reveal unique patterns and behaviors. This section examines these evolutionary trends for star-forming and quenched galaxies in voids, spanning a range of redshifts \(z \leq 2\) in Figures 3 and 4. Each plot includes two lines: a blue line representing star-forming galaxies and a red line for quenched void galaxies, which indicate the median values of each parameter. The shaded regions around each curve represent the median absolute deviation (MAD) of the values within each redshift bin, providing a robust estimate of variability. 

The first row of Figure 3 shows star-forming and quenched galaxies, which display a general increase in specific star formation rate (sSFR) as redshift rises across all stellar mass categories. This trend aligns with prior research on galaxy evolution, suggesting that galaxies had higher star formation rates in the early universe than today (\cite{speagle2014highly}). The plot's right side (High-mass) shows a slightly steeper increase in sSFR with redshift than the lower-mass bins (left side). This could indicate that high-mass galaxies had a more rapid star formation history, quenching faster and earlier due to more effective quenching mechanisms (\cite{peng2010mass}). Also, quenched galaxies in our sample in all mass ranges in low redshift reduce their sSFR more rapidly than star-forming. 
The second row shows the evolution of star formation rate (SFR) as a function of redshift across different stellar mass ranges for star-forming and quenched galaxies. In the stellar mass range \([10^8 - 10^{9}] M_\odot\), the SFR of quenched galaxies gradually declines as redshift decreases, especially around \(z=0.5\). Higher stellar mass range \([10^9 - 10^{10.5}] M_\odot\), the plot shows a more noticeable decrease in SFR for quenched void galaxies as redshift declines, especially after \(z=1\). At the same time, star-forming galaxies maintain a stable SFR trend again across redshift. At redshift \(>1\), quenched galaxies tend to have SFR values comparable to or even slightly higher than star-forming galaxies across all mass bins, indicating a delayed quenching effect in void environments at earlier times. Quenched galaxies have lower recent star formation histories (SFHs) and higher early SFHs. The elevated SFRs at early epochs suggest that quenched galaxies were once highly active in forming stars but eventually heat the cool gas of the interstellar medium by black hole thermal feedback (\cite{man2018star}; \cite{zinger2020ejective}) or gas outflows driven by stellar winds or supernova explosions that mainly acts on low-mass galaxies that are stopping further star formation activity (\cite{dekel2003feedback}). 
The third row of Figure 3 shows the gas mass (\(M_\text{gas}\)) within twice the half-stellar-mass radius (\(2 \times r_{1/2,*}\)) for star-forming and quenched galaxies across different mass bins plotted against redshift. In the low-mass range \([10^8 - 10^{9}] M_\odot\), star-forming and quenched galaxies exhibit similar gas masses across all redshifts. However, quenched galaxies in this mass range at low redshifts likely possess lower-density gas that cannot effectively support star formation, thus leading to their quenched state. In the high-mass range \([10^9 - 10^{10.5}] M_\odot\), quenched galaxies exhibit a slightly larger gas mass than star-forming galaxies at high redshifts. However, these quenched galaxies experience a marked reduction in gas within twice the half-mass radius at lower redshifts. This depletion may result from stronger feedback mechanisms, particularly stellar feedback( stellar winds or supernovas), which can efficiently expel gas in high-mass galaxies.

For star-forming and quenched void galaxies, a galaxy is defined as gas-rich if its gas fraction fulfills the criterion :\(gas   fraction \equiv M_{\text{gas}}/(M_{\text{gas}} + M_{\text{star}}) > 0.5\) (\cite{stewart2009gas}). 
For stellar mass ranges \([10^8 - 10^{9.5}] \, M_\odot\), our void galaxies are gas-rich across all redshifts. However, for higher stellar masses \([10^{9.5} - 10^{10.5}] \, M_\odot\), they remain gas-rich approximately at \(z > 0.5\), with a pronounced decrease in gas richness at lower redshifts, especially for quenched void galaxies.
Furthermore, galaxies at high redshifts show significantly higher gas fractions than those at low redshifts, consistent with the expectation that earlier epochs were characterized by abundant gas supplies (\cite{tacconi2020evolution}) and Star-forming galaxies maintain slightly higher gas fractions across all redshifts and mass bins than quenched galaxies. 
 
In TNG illustris simulation, a black hole is seeded in any friends-of-friends (FoF) halo whose mass exceeds \(7.3 \times 10^{10} \, M_\odot\) (if the halo does not already possess a BH), with an initial seed mass of \(6.2 \times 10^6 \, M_\odot\).Once seeded, BHs can grow by accreting gas from their close vicinity, the ‘feedback region’, which also defines the gas cells into which feedback energy is deposited. The size of the feedback region in the \(z = 0\) snapshots is of the order of a few kiloparsecs: \(\sim 3.6 \, \mathrm{kpc}\) in TNG300 on average(\cite{zinger2020ejective}). In addition, BH–BH mergers occur whenever one black hole is within the ‘feedback region’ of the other(\cite{sijacki2007unified}). 

In the first row of Figure 4, we explore the evolution of the median supermassive black hole(SMBH) mass (\(M_\text{BH}\)) across different stellar mass bins within twice the half-stellar-mass radius. In both star-forming and quenched void galaxies' mass ranges  \([10^8 - 10^{9.5}] M_\odot\) supernovae feedback limits the amount of available gas for black hole(BH) accretion, keeping the black hole mass close to the initial seed mass (though the BH can still grow by BH–BH mergers). As the galaxy grows in mass and its potential well deepens, the effectiveness of supernovae feedback diminishes. The black hole grows quickly to higher masses in low redshift, concentrating at roughly \(M_\text{BH} \sim 10^{8} \, M_\odot\)(\cite{pillepich2018simulating}). In this evolution, SMBH of star-forming void galaxies exhibits a slight increase in higher stellar mass, particularly within the mass range of  \([10^{9.5}- 10^{10}] M_\odot\).

In the second row, we show the evolution of the instantaneous black hole median accretion rate (\(\dot{M}_\text{BH}\)) of our void galaxies. The plot indicates that quenched galaxies rapidly decrease their black hole accretion rates at low redshifts in all mass ranges, while star formation enhances them, particularly in low-mass galaxies. Also, the evolution of black hole accretion rate (\(\dot{M}_\text{BH}\))  and SFR of starforming and quenched void galaxies are the same. The relation between SMBH accretion rate and SFR enhancements studied by \cite{byrne2023interacting}, demonstrated that galaxies in thermal mode accretion rate, on average, have both accretion rate and star formation rate enhancements, which accretion rate enhancements persist for up to two Gyrs after coalescence and are significantly longer lived than SFR enhancements, which only persist for \(\sim 500 \, \text{Myrs} \, \text{post-merger}\). This relation was also studied by \cite{sijacki2015illustris} that Gas-rich major mergers, which do bring copious amounts of gas to the center, are responsible for the star formation rate – BH accretion rate connection but relative timing offsets (\cite{wild2010timing}) Moreover, it concluded that in the case of large scale gas inflows and minor mergers star formation events might not be followed by black hole accretion because of gas consumption, expulsion or a residual angular momentum barrier. Because the number of mergers in void galaxies is higher at later times than in non-void galaxies (\cite{rodriguez2024evolutionary}), and void galaxies exhibit a higher fraction of accreted mass at low redshifts(\cite{dominguez2023galaxies}), may our star-forming galaxies experience higher merger rate and mergers in later times than quenched void galaxies which enhances the SFR and accretion rate of SMBH.   
Finally, the third row highlights the evolution of median dark matter mass (\(M_\text{DM}\)) across redshifts, showing that quenched galaxies generally evolve in a relatively higher gravitational potential well (i.e.higher dark matter haloes) than star-forming galaxies, which may impact their galactic properties. This trend is observed across all redshifts and stellar mass bins, and the discrepancy increases in low redshifts. 

\begin{figure*}[t] 
    \centering
    \includegraphics[width=\textwidth, keepaspectratio]{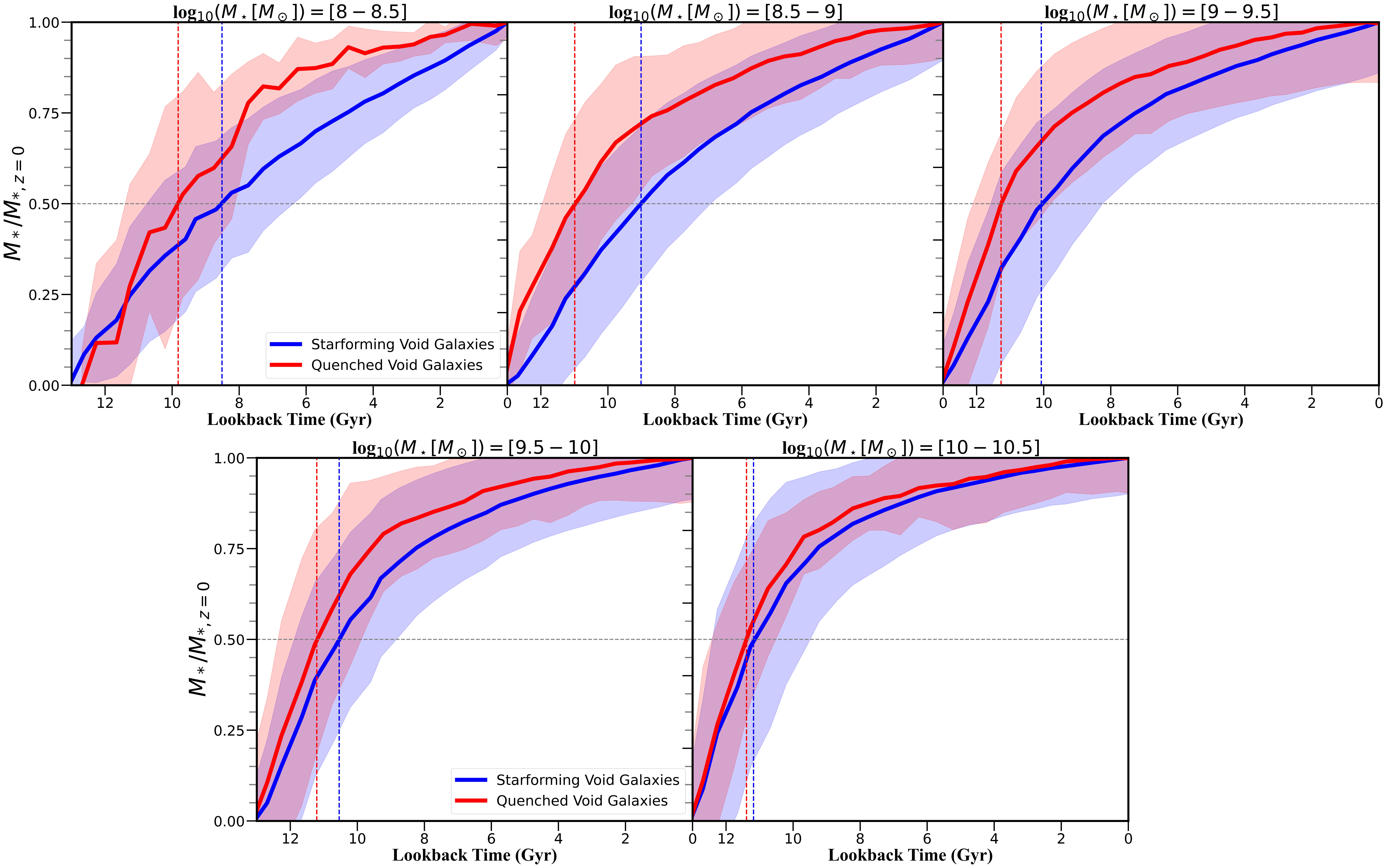}
 \caption{Stellar mass assembly histories of star-forming (blue) and quenched (red) galaxies across five stellar mass bins as a function of lookback time(Gyr). The normalized stellar mass fraction is shown with solid lines representing medians and shaded regions indicating the median absolute deviation (MAD). Vertical dashed lines mark the formation redshift (\(t_{\text{form}}\)), where 50\% of the final stellar mass is assembled. Quenched galaxies consistently form earlier than star-forming, with the largest time difference observed in low-mass galaxies.}

    \label{fig:stellar_mass_5comparison}
\end{figure*}

\subsection{Stellar Mass Assembly Histories of Void Galaxies}

This section analyses the evolution of star-forming and quenched void galaxies by examining their stellar mass assembly histories. To quantify the formation time of galaxies, we define \(t_{\text{form}}\) as the lookback time at which the main progenitor has assembled 50\% of its final stellar mass at \(z=0\). To calculate \(t_{\text{form}}\), we use a normalized stellar mass fraction defined as \(M_*/M_{*, z=0}\), where \(M_*(z)\) is the stellar mass at a given redshift.

Figure 5 illustrates the growth in stellar mass relative to the final stellar mass at  \(z=0\). For each lookback time bin, we compute the median values of this normalized mass fraction to track galaxies' stellar mass growth over time. The shaded regions around the curves represent the median absolute deviation (MAD) within each redshift bin, providing a robust measure of variability in the data. The vertical dashed blue and red lines indicate the lookback time at which star-forming and quenched void galaxies have assembled 50\% of their final stellar mass. These lines correspond to the \(t_{\text{form}}\) values for each population. 
 This plot shows that higher mass assembles their mass earlier than low mass in void environments, where quenched void galaxies form 50\%  of their mass in all stellar mass bins earlier than star-forming void galaxies. This discrepancy is more significant in low-mass ranges and decreases when we examine higher-mass. Earlier formation and rapid evolution of quenched galaxies may be the reason for their higher SFR and accretion rates of BH in high redshift. Table 1 represents the median lookback time \(t_{\text{form}}\) and 75th - 25th Percentiles of star-forming and quenched void galaxies in 5 mass bins.   
As we showed, quenched galaxies, due to their early formation and the fact that they reside in heavier dark matter halos, may consume their gas more rapidly and exhibit higher star formation at high redshift. This trend, influenced by feedback mechanisms, declined at low redshift.

\renewcommand{\arraystretch}{1.5} 
\begin{table}[h!]
\centering
\begin{tabular}{|c|c|c|}
\hline
& \multicolumn{2}{c|}{\textbf{Lookback Time [Gyr]}} \\
{\textbf{log$_{10}$[M$_{\odot}$]}}& \textbf{Star-forming Galaxies} & \textbf{Quenched Galaxies} \\
\hline
8.0 - 8.5 & $8.51 \, \substack{+0.33 \\ -0.41}$ & $9.82 \, \substack{+0.51 \\ -0.25}$ \\
8.5 - 9.0 & $9.01 \, \substack{+0.26 \\ -0.44}$ & $10.99 \, \substack{+0.37 \\ -0.43}$ \\
9.0 - 9.5 & $10.07 \, \substack{+0.26 \\ -0.59}$ & $11.27 \, \substack{+0.54 \\ -0.11}$ \\
9.5 - 10.0 & $10.54 \, \substack{+0.28 \\ -0.48}$ & $11.21 \, \substack{+0.16 \\ -0.59}$ \\
10.0 - 10.5 & $11.18 \, \substack{+0.18 \\ -0.62}$ & $11.39 \, \substack{+0.49 \\ -0.26}$ \\
\hline
\end{tabular}
\caption{The median lookback times of \( t_{\text{form}, *} \) [Gyr], the galaxy formation time, is defined as the lookback time at which the main progenitor has assembled 50\% of its final stellar mass (at \( z = 0 \)) for star-forming and quenched void galaxies. The supra-index and under-index values represent the difference between the median and the 75th and 25th percentiles, respectively.}

\end{table}
In addition to the reasons previously mentioned factors contributing to the suppression of SFR of quenched galaxies at low redshifts and uniformly
SFR over time for star-forming void galaxies, galaxy merger rates can significantly influence their star formation rates(\cite{davies2015galaxy};\cite{patton2016galaxy}; \cite{garduno2021galaxy}; \cite{moreno2021spatially}). By focusing on void galaxies, the level of star formation activity seems to be modulated by other mechanisms, such as the lack of replenishing gas and/or mergers and interactions (\cite{rosas2022revealing}), our research aims to better understand how galaxy merger rates affect star formation in these unique environments. 

\begin{figure}[t] 
    \centering
    \includegraphics[width=0.5\textwidth]{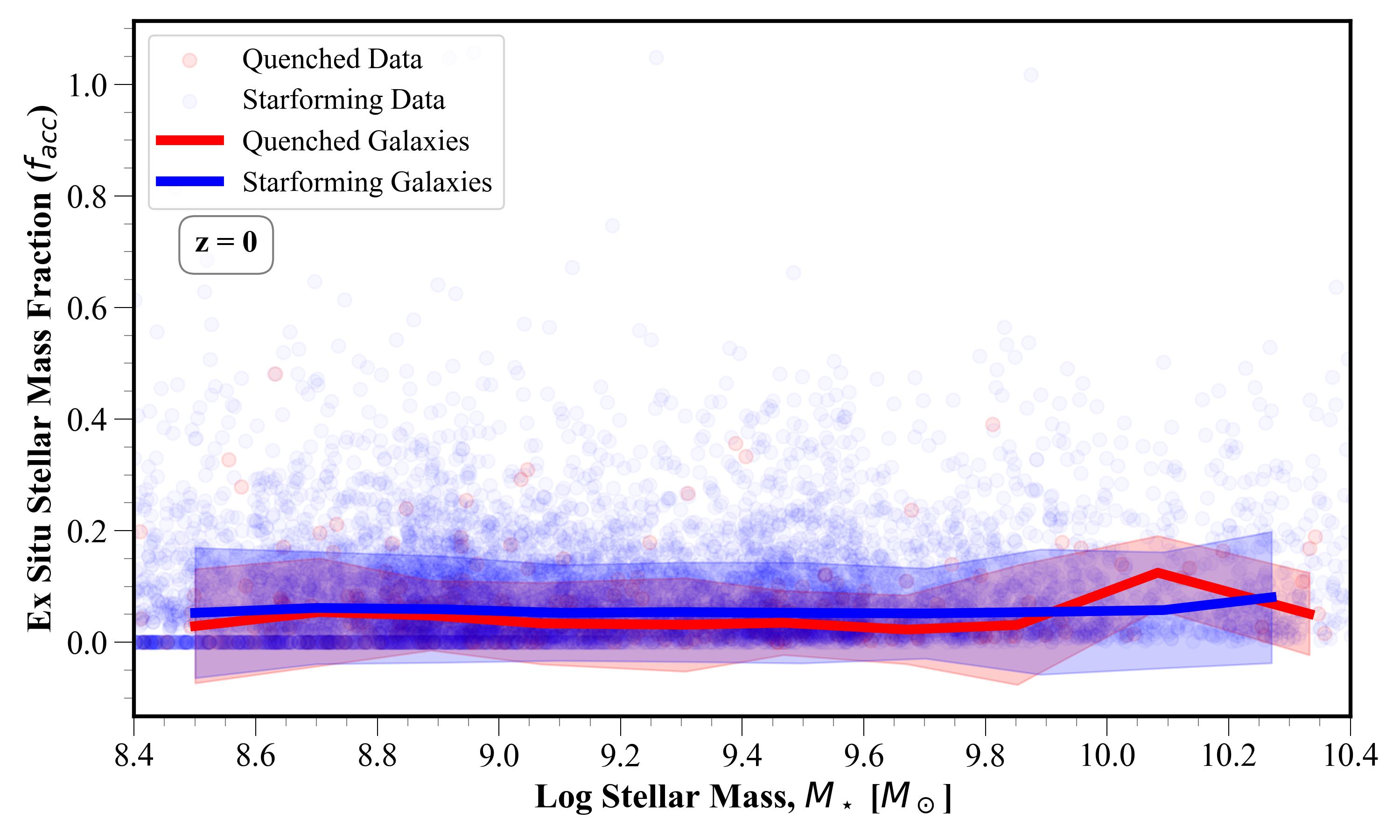}
    \caption{The median ex-situ stellar mass fraction \( f_{\text{acc}} \) as a function of
stellar mass \( M_{\ast} \) is shown for void galaxies since \( z=5 \) (\( \approx 12.6 \, \text{Gyr} \)). Star-forming and quenched void galaxies are shown in blue and red colors. The shaded region represents the standard deviation around the median with the \( 1 \sigma \) scatter, and the scattered dots represent data points for the two populations.}

    \label{fig:UVJ_diagram} 
\end{figure}

\subsection{Merger Histories of Void Galaxies}
Understanding the importance of mergers in the context of galaxy formation history requires investigating the rates of galaxy mergers across various cosmic periods (\cite{conselice2009structures}). As voids are environments with a low density of galaxies, it could be expected that the rate of interactions of galaxies is lower than in denser environments (\cite{van2011cosmic}). However, simulations predict that the number of galaxy mergers does not depend on the environment, and galaxies within voids have undergone more recent mergers compared to galaxies in other environments, indicating a different assembly rate (\cite{rosas2022revealing}; \cite{rodriguez2024evolutionary}). Additionally, galaxy interactions are particularly effective at stimulating substantial star formation in low- to moderate-density environments (\cite{sol2006effects}; \cite{das2021galaxy}). Furthermore, wet mergers are more prevalent in under-dense regions (\cite{lin2010wet}). Therefore, we analyze the merger rate of star-forming and quenched galaxies in void environments in various cosmic periods.

For our purpose, we use the publicly available merger history catalogue of \cite{rodriguez2017role} and \cite{eisert2023ergo} which contains information regarding galaxy mergers and the ex-situ (accretion of stars, where stars are born in other smaller galaxies) and in-situ (stars are formed within the galaxy from infalling cold gas) stellar formation processes. The time of the merger is defined as the time when the corresponding merger tree branches join. Also, for deep analysis of the effect of mergers on star-forming and quenched galaxies, mergers are split into three categories: major (stellar mass ratio > 1/4), minor (stellar mass ratio between 1/10 and 1/4), and all (any stellar mass ratio) mergers. The mass ratio is always based on the stellar masses of the two merging galaxies when the secondary reaches its maximum stellar mass (\cite{rodriguez2015merger}). 

The primary factor we use to quantify the importance of merging history is the ex-situ stellar mass fraction, symbolized by \( f_{\text{acc}} \), which measures the average amount of stellar mass that
a galaxy accretes per unit time through mergers with other galaxies. It is essential to emphasize that \( f_{\text{acc}} \) does not directly measure the history of merging; rather, it evaluates the significance of dry merging in relation to dissipative processes like in situ star formation(\cite{oser2010two}). 

Figure 6 presents the ex-situ stellar mass fractions of galaxies since $z=5$  at redshift $z=0$ as a function of their stellar mass $M_*$, using data derived from the TNG300 simulation. The plot showcases median values of the ex-situ stellar mass fractions for quenched and star-forming galaxies, represented by red and blue lines, respectively. Individual data points scattered across the plot illustrate each galaxy type's range and distribution of values, and the shaded region represents the standard deviation around the median with the $1 \sigma$ scatter. As discussed in \cite{rodriguez2017role} at any fixed stellar mass –\( f_{\text{acc}} \) is negatively correlated with gas-rich mergers, and higher \( f_{\text{acc}} \) is associated with a more significant number of massive, recent, and dry mergers that are strong for massive galaxies ( \(M_\text{*} > 10^{11} \, M_\odot\)). Therefore, observed ex-situ stellar mass fractions of our void galaxies below 0.1 across a broad range of galaxy stellar masses suggest a predominant role of in-situ star formation in these galaxies' development that indicates that most stars in these void galaxies were formed from accreted gas within the galaxy and gas-rich mergers are dominated in our void galaxies.

\begin{figure*}[t] 
    \centering
    \includegraphics[width=\textwidth, keepaspectratio]{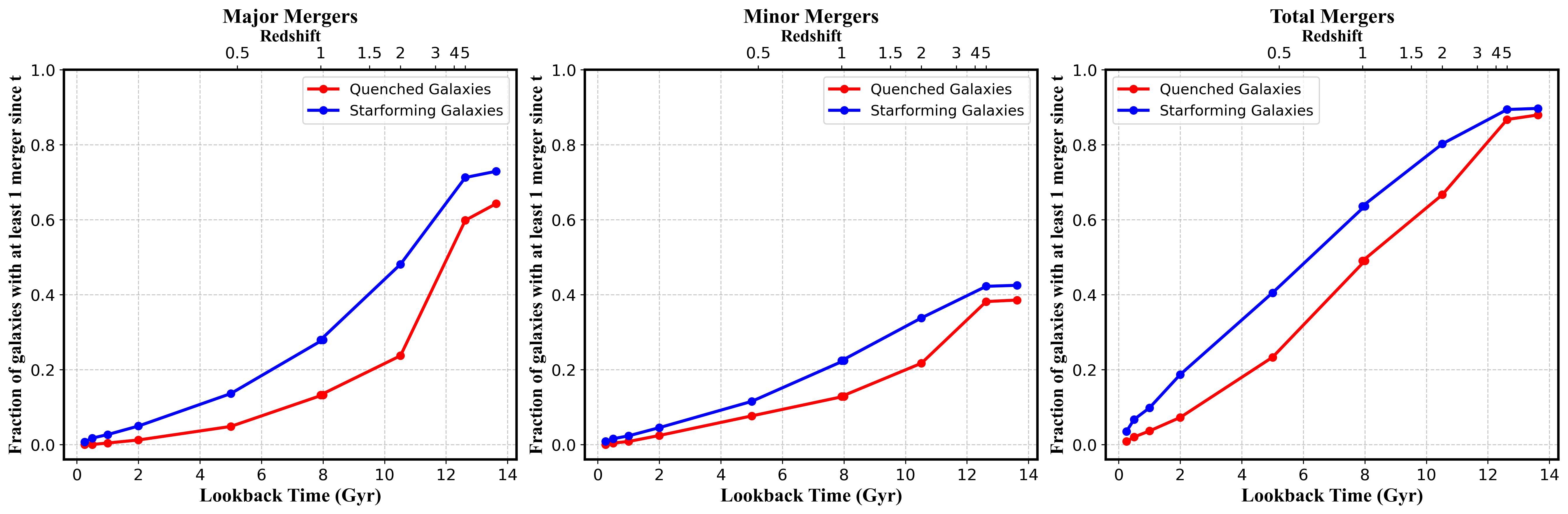}
    \caption{fraction of galaxies that have undergone at least one merger in different periods (lookback time
measured from $z=0$) and different stellar mass ratios: major minor,  and total mergers for star-forming and quenched galaxies in the void that blue lines and red lines indicate the star-forming and quenched fraction of mergers respectively. Dots represent the
exact periods for which the measurements are available.}
    \label{fig:stellar_mass_6comparison}
\end{figure*}

\begin{figure*}[t] 
    \centering
    \includegraphics[width=\textwidth, keepaspectratio]{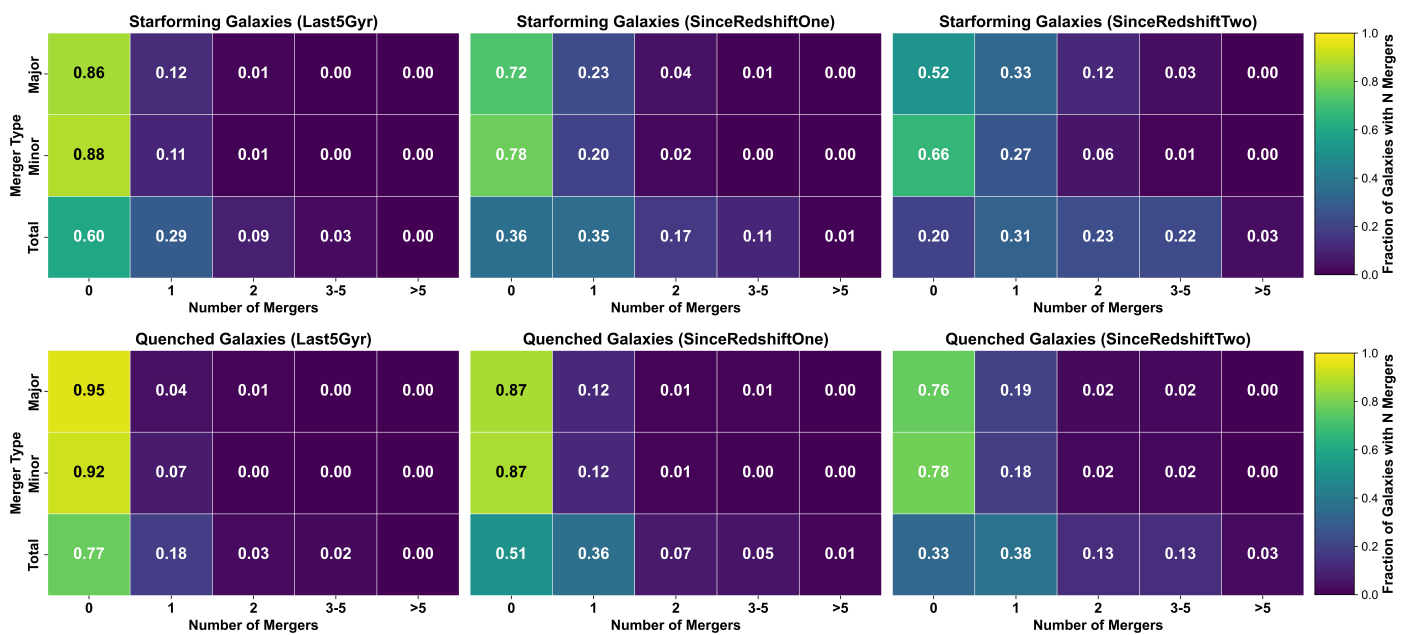}
   \caption{Fraction of galaxies undergoing different types of mergers for three defined time periods.
Left: Last 5 Gyr ($z \approx 0.5$). Middle: Since $z=1$ (Lookback time $\approx 7.92$ Gyr). Right: Since $z=2$ (Lookback time $\approx 10.51$ Gyr). The color denotes the fraction of galaxies undergoing a certain number of mergers.}

    \label{fig:stellar_mass_7comparison}
\end{figure*}

\subsubsection{Merger statistics}
Figure 7 provides a detailed visualisation of the fraction of star-forming and quenched galaxies undergoing at least one merger over various lookback times (Gyr) from  $z=0$, segmented by major, minor, and total merger types(any stellar mass ratio). The data is depicted by blue lines for star-forming galaxies and red lines for quenched galaxies; dots represent the exact periods for which the measurements are available. Both galaxy types experience an increase in merger fractions as we track back to higher redshifts. The plot shows that star-forming void galaxies have a higher merger activity (major, minor, and mergers in any mass ratio) than quenched galaxies over the given lookback time. 

In Figure 8, we explore the fractions of star-forming and quenched galaxies that have experienced varying numbers of mergers across three key epochs in cosmic history. These periods include the last 5 Gyr since \( z \approx 0.5 \), since \( z = 1 \) (lookback time approximately 7.98 Gyr), and since \( z = 2 \) (lookback time approximately 10.51 Gyr). The heatmap's color coding effectively illustrates the fraction of galaxies within each category that underwent a certain number of mergers, providing a detailed visual representation of merger activity over time.
In analyzing merger activities over the past several billion years, distinct patterns emerge between star-forming and quenched galaxies, illustrating their divergent evolutionary paths. During the last 5 Gyr (since \(z \approx 0.5\)), Star-forming void galaxies have a lower fraction of zero mergers (no mergers) across all major, minor, and total merger types. Additionally, star-forming void galaxies have a higher fraction of one merger than quenched void galaxies in all merger types. However, the fraction of more than one merger is approximately similar between the two populations. As we extend our view to about 7.98 Gyr ago (since \(z = 1\)), star-forming void galaxies reduced no-total merger to 36\%, exhibited higher fraction of one minor and major mergers 23\% and 20\%, respectively, and increased two total mergers increased to 17\%.
By the epoch since \(z = 2\) (approximately 10.51 Gyr ago), star-forming galaxies show the most remarkable diversity in their merger activities, with a significant reduction in no-total merger instances to 20\% and an increase in the fraction of major mergers to 33\% and 12\% for one and two mergers, respectively, compared to 19\% and 0.02\% merger fractions for quenched galaxies and a notable increase in those experiencing more than two total mergers. These observations reveal that star-forming void galaxies exhibit more active merger histories, with lower fractions of no mergers and more frequent, minor, and major mergers, particularly since redshift two and higher levels of recent merger activity than quenched void galaxies. 
\begin{table*}[htbp]
\centering
\caption{Detailed Merger History Statistics for Star-forming and Quenched Galaxies. Numbers represent median  values of the last major, minor, and total mergers across different mass bins in lookback times. The supra index and under index median values represent the difference between the median and 75th and 25th percentiles }
\label{tab:merger_history}
\scalebox{1.1}{ 
\begin{tabular}{|c|c|c|c|c|c|}
\hline
\textbf{Galaxy Type / Mass Bin Log10} & \textbf{[8.0 - 8.5]$M_*$} & \textbf{[8.5 - 9.0]$M_*$} & \textbf{[9.0 - 9.5]$M_*$} & \textbf{[9.5 - 10.0]$M_*$} & \textbf{[10.0 - 10.5]$M_*$} \\
\hline
& & & & & \\
Star-forming Minor Mergers& \( 0.42 \substack{+0.40 \\ -0.23} \) & \( 0.64 \substack{+0.57 \\ -0.37} \) & \( 0.92 \substack{+0.68 \\ -0.48} \) & \( 1.21 \substack{+0.79 \\ -0.61} \) & \( 1.41 \substack{+1.17 \\ -0.77} \) \\
& & & & & \\
Star-forming Major Mergers& \( 1.07 \substack{+0.79 \\ -0.61} \) & \( 1.07 \substack{+0.83 \\ -0.59} \) & \( 1.41 \substack{+0.91 \\ -0.71} \) & \( 1.74 \substack{+1.27 \\ -0.92} \) & \( 1.81 \substack{+1.67 \\ -1.12} \) \\
& & & & & \\
Star-forming Total Mergers& \( 0.33 \substack{+0.31 \\ -0.17} \) & \( 0.50 \substack{+0.50 \\ -0.30} \) & \( 0.50 \substack{+0.45 \\ -0.30} \) & \( 0.38 \substack{+0.44 \\ -0.23} \) & \( 0.21 \substack{+0.25 \\ -0.13} \) \\
\hline
& & & & & \\
Quenched Minor Mergers& \( 0.84 \substack{+0.26 \\ -0.26} \) & \( 1.27 \substack{+0.65 \\ -0.80} \) & \( 2.00 \substack{+0.90 \\ -0.59} \) & \( 2.32 \substack{+1.17 \\ -1.53} \) & \( 0.38 \substack{+1.44 \\ -0.07} \) \\
& & & & & \\
Quenched Major Mergers& \( 2.87 \substack{+1.38 \\ -1.41} \) & \( 2.00 \substack{+1.01 \\ -0.97} \) & \( 2.44 \substack{+1.05 \\ -1.19} \) & \( 3.01 \substack{+0.70 \\ -1.13} \) & \( 1.07 \substack{+0.79 \\ -0.61} \) \\
& & & & & \\
Quenched Total Mergers& \( 0.84 \substack{+0.26 \\ -0.26} \) & \( 0.69 \substack{+0.88 \\ -0.35} \) & \( 1.21 \substack{+0.79 \\ -0.51} \) & \( 0.79 \substack{+0.84 \\ -0.57} \) & \( 0.23 \substack{+0.25 \\ -0.17} \) \\
\hline
\end{tabular}
}
\end{table*}

Table 2 details the median lookback time (Gyr) of the last major, minor, and total merger(any stellar mass ratio )  for star-forming and quenched void galaxies for five stellar mass bins. Notably, the supra-index and sub-index accompanying the median values indicate the differences between the median and the 75th and 25th percentiles, respectively. For star-forming void galaxies, the last minor mergers lookback time statistically increases around \( 0.42 \substack{+0.40 \\ -0.23} \)  Gyr to \( 1.41 \substack{+1.71 \\ -0.77} \) Gyr with increased stellar mass bins, indicating that higher mass has earlier last minor mergers. The same pattern was observed for the last major mergers; the more massive bins experienced their last major mergers earlier. For quenched void galaxies, we have access to reliable last mergers data in mass ranges  \([10^{8.5}- 10^{10}] M_\odot\),  exhibit a similar pattern for both minor and major mergers with star-forming galaxies. 
Table 2 demonstrates that star-forming void galaxies exhibit \textbf{more recent major mergers} compared to quenched galaxies, with median lookback times around 1.07 Gyr for low-mass bins and 1.74 Gyr for higher masses, in contrast, quenched galaxies' major mergers occurred earlier, with higher median redshifts (2.87 to 3.01 Gyr). Star-forming void galaxies also exhibit \textbf{more recent minor mergers} with median lookback times around 0.42 Gyr for low-mass bins and 1.21 Gyr for higher masses. Overall, the table confirms that star-forming galaxies consistently show lower median lookback time for all merger types, emphasizing their more recent merger activity than quenched galaxies, demonstrating earlier merger histories.

Figure 9 shows a heatmap that visualizes the percentage of star-forming and quenched void galaxies undergoing minor and major mergers across five mass bins and epochs: the last 12.5 Gyr, the last 10.5 Gyr, 8 Gyr, 5 Gyr, and the last 2 Gyr) in the TNG300 simulation. The first two columns represent minor mergers for star-forming and quenched galaxies, respectively, while the third and fourth columns represent major mergers for the same populations. Each row shows non-merger, at least one merger, and more than one for minor and major mergers of galaxies across cosmic time. We have to emphasize that our minor merger data are higher in mass ranges\([10^{8}- 10^{10}] M_\odot\) for quenched void galaxies, and our results are more reliable for these mass ranges. The heatmap shows the minor and major merger rates evolve with redshift, and that their percentages consistently decrease from $z=5$ to the present (\cite{lotz2011major}) that, indicating that these two populations in all mass ranges have higher percentages of mergers in earlier times of evolution.  
\begin{figure*}[t] 
    \centering
    \includegraphics[width=\textwidth, keepaspectratio]{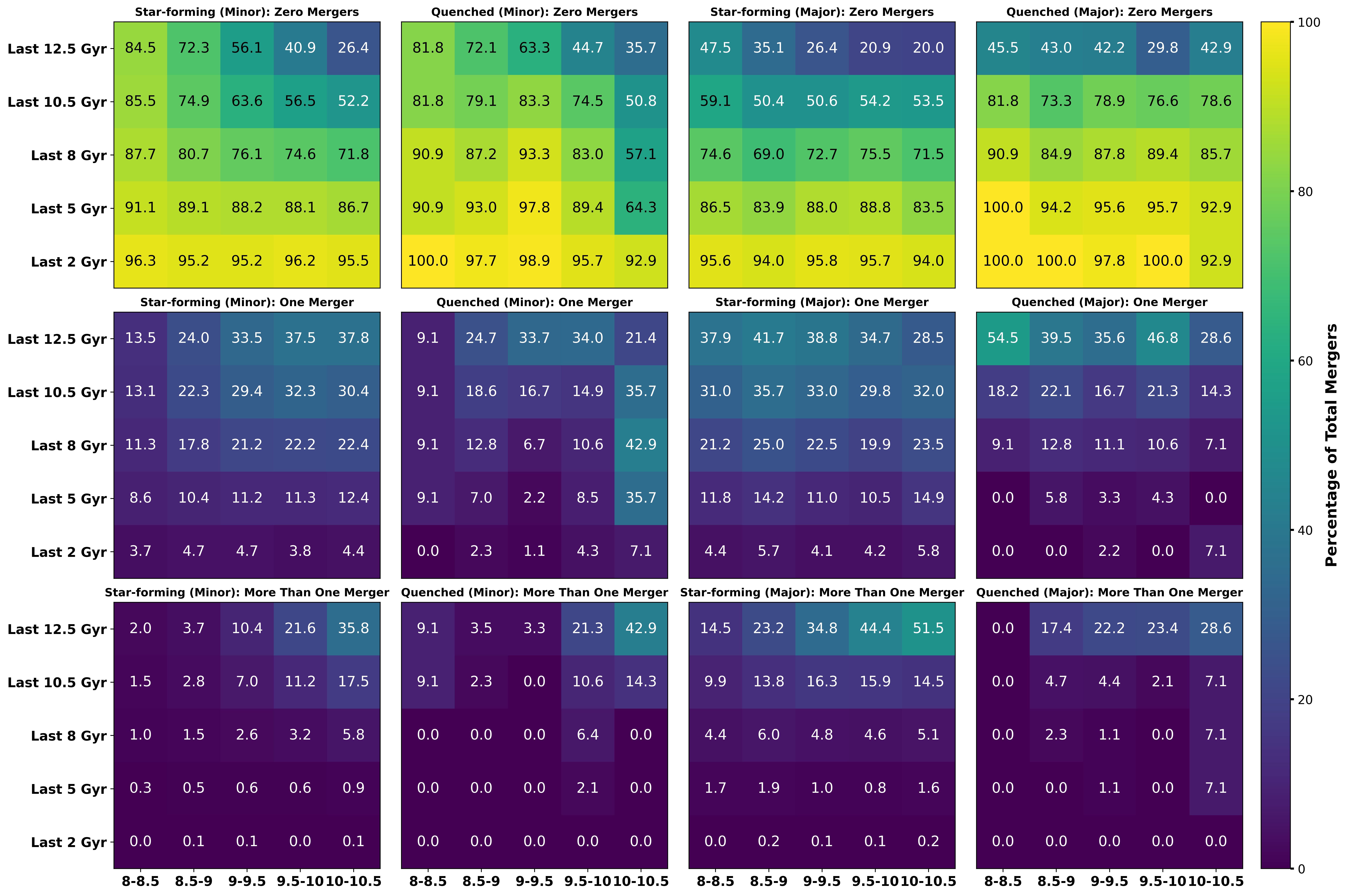}
    \caption{Percentage of minor (columns 1 and 2) and major mergers (columns 3 and 4) for total mergers the last 12.5 Gyr, 10.5 Gyr, 8 Gyr, 5 Gyr, and the last 2Gyr, for star-forming and quenched void galaxies in TNG300. The five mass bins in $\log_{10} M_{\odot}$ are indicated along the x-axis, and the color represents the percentage of galaxies that underwent minor or major mergers in each mass bin relative to the total galaxies in each bin. }
    \label{fig:stellar_mass_10comparison}
\end{figure*}

In the comparison between quenched and star-forming void galaxies, the trends in mergers reveal distinct evolutionary pathways. For \textbf{one minor merger}, the percentages are approximately similar between the two populations over the last 12.5 Gyr. However, in some mass bins \([10^{9.5}- 10^{10}] M_\odot\), star-forming galaxies show slightly higher percentages, suggesting that minor mergers have been relatively common for both populations early on. This trend looks noticeably differently in recent epochs, particularly over the last 10.5 Gyr and especially the last 8 Gyr, where star-forming galaxies exhibit a higher fraction of one minor merger event. This indicates that star-forming galaxies have undergone more recent minor interactions, which have been evident over the last 5 Gyr. For \textbf{one major merger}, quenched galaxies dominated in the last 12.5 Gyr for specific mass bins; however, this trend diminishes in more recent epochs, particularly over the last 8 Gyr, where star-forming galaxies exhibit a higher proportion of galaxies with one major merger. This implies that in their recent evolution, major mergers have become more significant for star-forming galaxies. 
Regarding \textbf{more than one merger}, star-forming galaxies consistently show significantly higher percentages than quenched galaxies, particularly for major mergers. This trend is prominent not only in the last 12.5 Gyr but also in recent times, indicating that multiple major mergers are a crucial factor in the evolution of star-forming galaxies. These repeated major mergers likely sustain or even enhance star formation by introducing fresh gas or triggering starburst activity, while quenched galaxies appear to have experienced fewer such interactions. 

To investigate the factors contributing to the stability of star formation rates (SFR) in star-forming galaxies and the declining SFR in quenched galaxies over the last 8 Gyr ( \(z \approx 1\))), we analyze the percentage of all mergers (any stellar mass ratio) in the last 8, 5, and 2 Gyr for these two populations, as shown in Figure 10. In the case of \textbf{one mergers}, star-forming galaxies exhibit a higher percentage of mergers in the last 5 Gyr, particularly in the mass range \([10^{8}- 10^{10}] M_\odot\). This trend becomes even more pronounced in the last 2 Gyr, where star-forming galaxies consistently show higher percentages across all mass bins than quenched galaxies. Additionally, for \textbf{multiple mergers}, star-forming galaxies display a strikingly higher and significantly more frequent occurrence of mergers over the last 8 Gyr. This discrepancy is especially notable in the higher mass bins, where the rates of multiple mergers for star-forming galaxies far exceed those of quenched galaxies. Interestingly, this trend persists in the last 5 Gyr and the last 2 Gyr, suggesting that star-forming galaxies have experienced a higher frequency of multiple mergers in recent times.

These findings indicate that the significantly higher rate of multiple major mergers over the last 12.5 Gyr, along with the intriguingly higher frequency of multiple mergers across all mass ratios in recent epochs, could play a pivotal role in maintaining the stable star formation rate (SFR) observed in star-forming galaxies. On the other hand, the rapid and earlier formation of quenched galaxies, coupled with their lower rate of recent mergers and major mergers throughout cosmic time, appears to be a key factor contributing to the decline in their SFR at low redshifts. This contrast highlights the critical influence of merger timing and frequency on the divergent evolutionary trajectories of star-forming and quenched galaxies in void environments.

\subsection{Gas Mass and SFR of Merger Galaxies}

To further analyze the impact of mergers on star-forming and quenched galaxies, we conducted a separate study comparing galaxies with no mergers to those with at least one merger.  Figure 11 compares the distributions of star formation rate (SFR), specific star formation rate (sSFR), star formation efficiency (\(\text{SFE} = \text{SFR}/M_{\text{gas}}\)), and a gas fraction (\(f_{\text{gas}}\)) for merger and non-merger groups of star-forming and quenched void galaxies. To ensure a detailed comparison, the galaxy samples are divided into two mass bins: low-mass (\(10^8-10^9\,M_\odot\)) and high-mass (\(10^{10}-10^{10.5}\,M_\odot\)). In each plot, red represents non-merger galaxies, while blue indicates galaxies with at least one merger. We perform a Kolmogorov-Smirnov (K–S) test on all groups to measure the significance of the difference. 

\begin{figure}[t] 
    \centering
    \includegraphics[width=1\columnwidth, keepaspectratio]{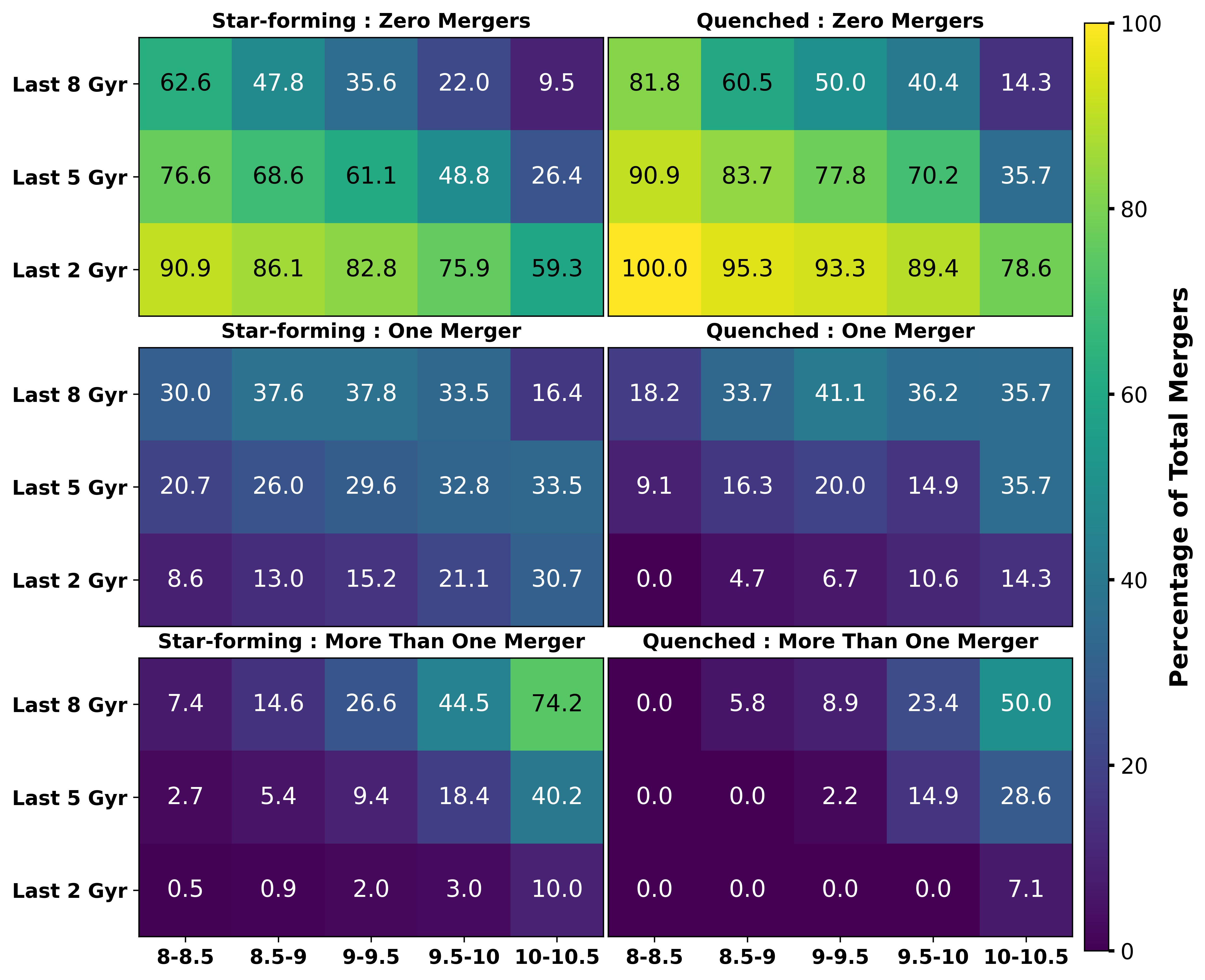} 
  \caption{Percentage of all mergers (any stellar mass ratio) for the last 8 Gyr, 5 Gyr, and 2 Gyr, for star-forming and quenched void galaxies in TNG300. The five mass bins in $\log_{10} M_{\odot}$ are indicated along the x-axis, and the color represents the percentage of galaxies that underwent all mergers in each mass bin relative to the total galaxies in each bin.}

    \label{fig:stellar_mass_11comparison}
\end{figure}

The analysis of low-mass quenched void galaxies (\(10^8-10^9\,M_\odot\)) demonstrates that mergers have a negligible impact on their physical properties, including star formation rate (SFR), specific star formation rate (sSFR), star formation efficiency (SFE), and gas fraction (\(f_{\text{gas}}\)). The Kolmogorov-Smirnov (KS) test results, with low KS statistics (e.g., \(D = 0.13-0.19\)) and high p-values (\(p > 0.49\)), indicate no statistically significant differences between the merger and non-merger populations for these parameters. Our analysis on high-mass quenched galaxies (\(10^{10}-10^{10.5}\,M_\odot\)) shows that in terms of star formation rate (SFR), the KS statistic is \(D = 0.48\) with a p-value of 0.03, indicating a marginally significant difference between merger and non-merger populations. However, the specific star formation rate (sSFR) exhibits minimal differences, with \(D = 0.25\) and a high p-value of 0.62. Similarly, the gas fraction (\(f_{\text{gas}}\)) shows some overlap, with \(D = 0.39\) and a p-value of 0.14, suggesting no strong statistical significance. But, star formation efficiency (SFE) demonstrates a more notable difference, with \(D = 0.51\) and a p-value of 0.02.

\begin{figure*}[t] 
    \centering
    
    \includegraphics[width=\textwidth, keepaspectratio]{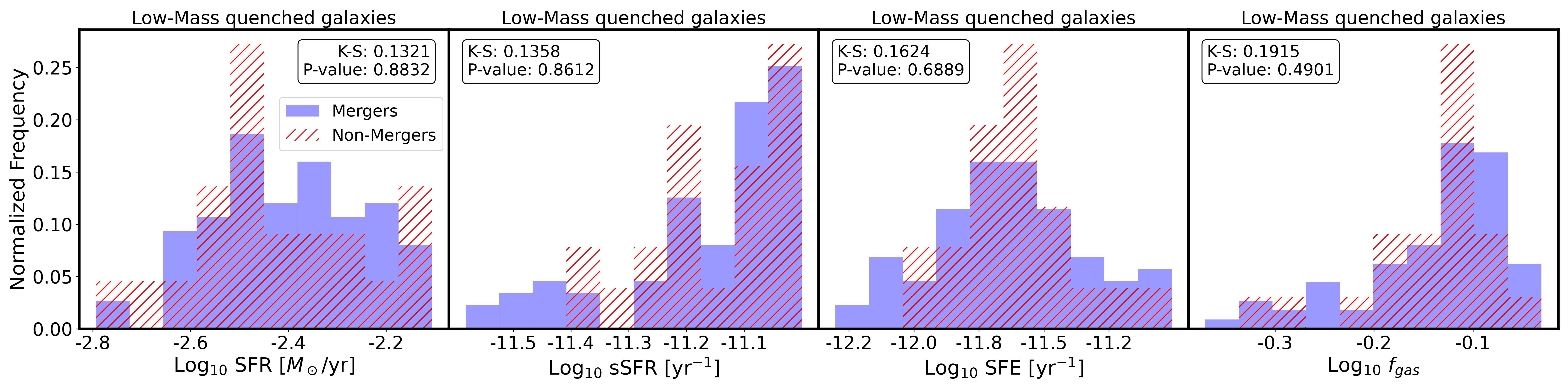}
    
    \vspace{1em} 
    
    \includegraphics[width=\textwidth, keepaspectratio]{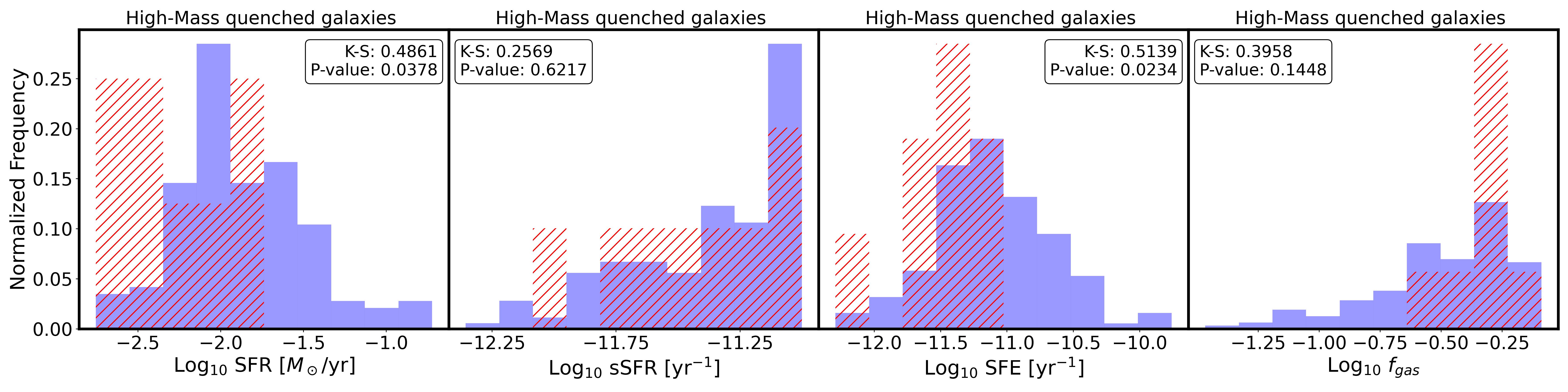}
    
    \vspace{1em} 
    
    \includegraphics[width=\textwidth, keepaspectratio]{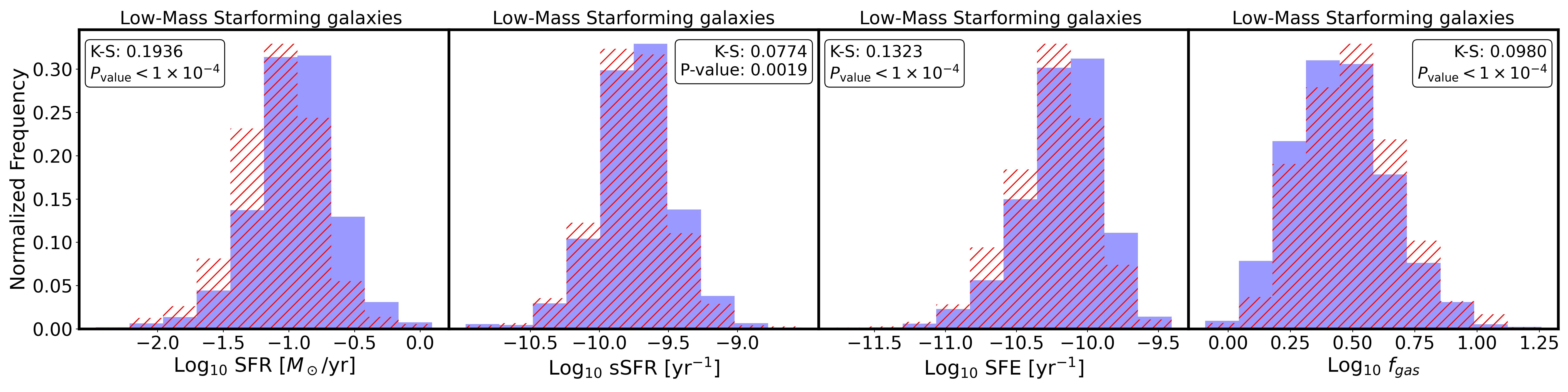}
    
    \vspace{1em} 
    
    \includegraphics[width=\textwidth, keepaspectratio]{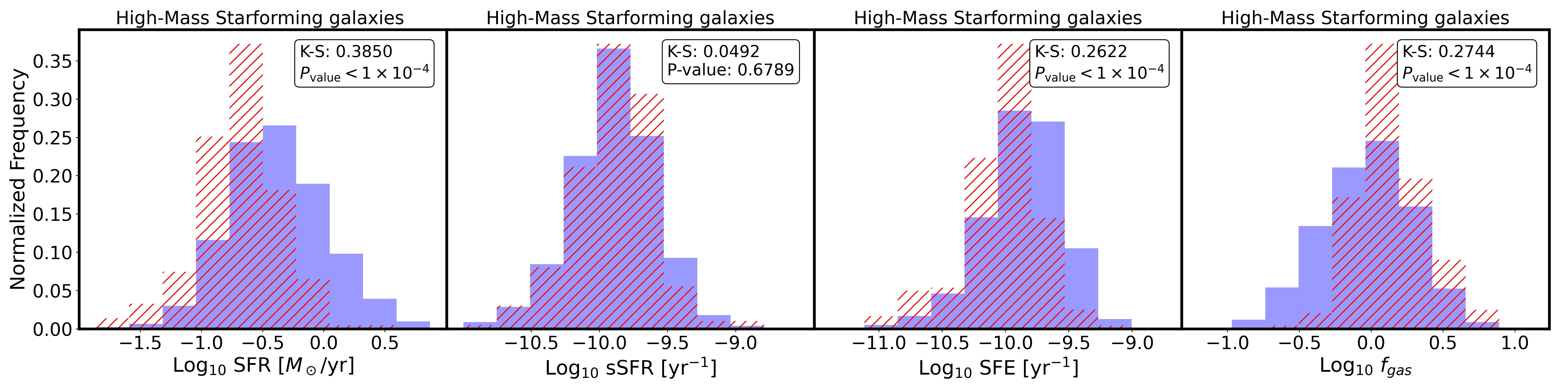}
    
    \caption{Histograms comparing the distributions of star formation rate (SFR), specific star formation rate (sSFR), star formation efficiency (SFE), and gas fraction ($f_\text{gas}$) for mergers (blue) and non-mergers (red) in void environments across different galaxy types and stellar mass ranges: low-mass galaxies ($10^8-10^9\ M_\odot$) and high-mass galaxies ($10^9-10^{10.5}\ M_\odot$). The Kolmogorov--Smirnov (KS) statistic and $p$-value are provided in each panel to quantify the statistical differences between the distributions.}
    \label{fig:01stellar_mass_comparison}
\end{figure*}

The analysis of low-mass (\(10^8-10^9\,M_\odot\)) and high-mass (\(10^9-10^{10.5}\,M_\odot\)) star-forming void galaxies reveals significant differences in their physical properties between merger and non-merger populations. For low-mass galaxies, the star formation rate (SFR) shows a notable difference with \(D = 0.19\) and \(p < 10^{-4}\), while specific star formation rate (sSFR) also exhibits variation with \(D = 0.07\) and \(p = 0.0019\). Similarly, star formation efficiency (SFE) and gas fraction (\(f_{\text{gas}}\)) follow the same trend, with \(D = 0.13\) and \(D = 0.09\), respectively, and \(p < 10^{-4}\) in both cases. For high-mass galaxies,  SFR displays a substantial discrepancy with \(D = 0.38\) and \(p < 10^{-4}\), while SFE and \(f_{\text{gas}}\) also exhibit significant differences with \(D = 0.26\) and \(D = 0.27\), respectively, and \(p < 10^{-4}\). In contrast, sSFR shows minimal variation between merger and non-merger populations, with \(D = 0.04\) and \(p = 0.67\), indicating no statistically significant difference. Our finding shows quenched void galaxies with at least one merge and non-merger galaxies have no significant differences in most properties except for marginal effects on SFR and SFE in high-mass quenched galaxies. In contrast, star-forming void galaxies with at least one merger significantly differ from non-merger galaxies in parameters such as SFR, SFE, sSFR, and \(f_{\text{gas}}\). However, for high-mass galaxies, the difference in sSFR between merger and non-merger populations is minimal.

As demonstrated, quenched galaxies formed earlier than star-forming galaxies, resulting in a significant depletion of their gas reservoirs for star formation at lower redshifts. Most mergers in quenched galaxies occurred earlier, limiting their impact at \( z = 0 \). However, more substantial gravitational potential enables gas accretion in high-mass quenched galaxies, leading to slightly higher star formation rates (SFR) than in non-merger quenched galaxies. In contrast, the later formation of star-forming void galaxies ensures they retain sufficient gas for sustained star formation. Additionally, the lack of environmental effects, such as ram pressure stripping, helps preserve their gas reservoirs. Star-forming galaxies with mergers benefit from gas inflows triggered by these events, enhancing their SFR, mainly through recent mergers that dynamically replenish central gas reservoirs.

In Table~\ref{tab:ks_minor_major_comparison_v3}, we present a statistical comparison of non-minor versus minor mergers and non-major versus major mergers in both star-forming and quenched galaxies, focusing on systems that have experienced at least one minor or major merger.
For \textbf{quenched galaxies}, the p-values across most parameters (SFR, sSFR, SFE, and $f_\text{gas}$) indicate statistically insignificant differences between non-minor and minor mergers and non-major and major mergers. This suggests that the impact of mergers on quenched galaxies is limited, likely because these systems have already ceased significant star formation, reducing their sensitivity to external influences like mergers. Notably, in high-mass quenched galaxies, the comparison between non-major and major mergers for SFR ($p$-value = 0.0319) suggests a weak but potentially meaningful distinction, possibly hinting at a minor resurgence of star formation in some cases.

For \textbf{star-forming galaxies}, the differences are far more pronounced. Low $p$-values, particularly for SFR, SFE, and $f_\text{gas}$, emphasise the significant influence of merger type on these galaxies' star formation activity and gas properties. For example, in high-mass star-forming galaxies, the KS statistic for SFR is 0.1696 with a $p$-value of $1.26\times10^{-35}$ for non-minor versus minor mergers and 0.1542 with a $p$-value of $4.15\times10^{-20}$ for non-major versus major mergers. These results suggest that both minor and major mergers in star-forming galaxies, especially at higher masses, drive considerable changes in star formation rates and efficiency, likely through processes like gas inflow, compression, and triggering of new star formation. The trends also indicate that major mergers have a stronger impact than minor mergers, consistent with the theoretical understanding of the larger gravitational disruptions caused by major events.

\begin{table*}[htbp]
    \centering
    \caption{Comparison of K-S Statistics and P-values for  Non-Minor vs. Minor Mergers and Non-Major vs. Major Mergers in Low-Mass ($10^8-10^9\ M_\odot$) and High-Mass ($10^9-10^{10.5}\ M_\odot$) of quenched and starforming void galaxies.}
    \begin{tabular}{|c|c|c|c|c|}
        \hline
        \textbf{Galaxy Type} & \textbf{Mass Range} & \textbf{Parameter} 
        & \textbf{Non-Minor vs. Minor} (KS, p-value) & \textbf{Non-Major vs. Major} (KS, p-value) \\
        \hline
        \multirow{4}{*}{Quenched} 
        & \multirow{4}{*}{ Low-Mass} 
        & SFR       & (0.1766, 0.5246) & (0.0987, 0.9520) \\
        & & sSFR      & (0.2925, 0.0604) & (0.1895, 0.3146) \\
        & & SFE       & (0.1874, 0.4509) & (0.2136, 0.1976) \\
        & & $f_{gas}$ & (0.1029, 0.9705) & (0.1110, 0.8933) \\
        \cline{2-5}
        & \multirow{4}{*}{High-Mass} 
        & SFR       & (0.2049, 0.0702) & (0.2431, 0.0319) \\
        & & sSFR      & (0.1502, 0.3193) & (0.0941, 0.8955) \\
        & & SFE       & (0.1341, 0.4529) & (0.2134, 0.0808) \\
        & & $f_{gas}$ & (0.1808, 0.1447) & (0.2005, 0.1174) \\
        \hline
        \multirow{4}{*}{Star-Forming} 
        & \multirow{4}{*}{Low-Mass} 
        & SFR       & (0.1315, $9.35\times10^{-11}$) & (0.0992, $2.13\times10^{-7}$) \\
        & & sSFR      & (0.0371, 0.2942) & (0.0579, 0.0084) \\
        & & SFE       & (0.0858, $7.93\times10^{-5}$) & (0.0719, 0.0004) \\
        & & $f_{gas}$ & (0.0829, 0.0002) & (0.0253, 0.6641) \\
        \cline{2-5}
        & \multirow{4}{*}{High-Mass} 
        & SFR       & (0.1696, $1.26\times10^{-35}$) & (0.1542, $4.15\times10^{-20}$) \\
        & & sSFR      & (0.0405, 0.0197) & (0.0461, 0.0346) \\
        & & SFE       & (0.1338, $2.58\times10^{-22}$) & (0.0979, $2.36\times10^{-8}$) \\
        & & $f_{gas}$ & (0.1349, $1.12\times10^{-22}$) & (0.0806, $8.43\times10^{-6}$) \\
        \hline
    \end{tabular}
    \label{tab:ks_minor_major_comparison_v3}
\end{table*}

\section{Discussion \& Conclusions }

This study utilizes the high-resolution TNG300 simulation from the IllustrisTNG (\cite{springel2018first}; \cite{pillepich2018first}) project galaxies with stellar masses (\(M_{\star} \geq 10^8\,M_{\odot}\)) to investigate the evolution histories of star-forming and quenched void galaxies and the impact of galaxy mergers on star formation rates (SFR) in void environments. This research represents the first statistical examination of merger rates—distinguishing between major and minor mergers—across different evolutionary stages of star-forming and quenched galaxies in voids, a largely unexplored study area. By integrating data from the "Merger History" catalog, it offers comprehensive insights into merger rates across cosmic time (\cite{rodriguez2017role}; \cite{eisert2023ergo}) and their critical role in shaping galaxy evolution in these isolated, low-density environments. 

First, we compared the main properties of star-forming and quenched void galaxies at  \(Z = 0\) within two times the half-mass stellar radius. Our findings indicate that quenched galaxies exhibit a lower gas number density across all ranges of stellar mass. Moreover, they possess a smaller gas mass compared to star-forming void galaxies, with this effect being more significant in the higher mass ranges of [\(10^9-10^{10.5}]\,M_\odot\). This could explain the lower star formation rate (SFR) and specific star formation rate (sSFR) observed in the present. We also compared the mass of supermassive black holes(\(M_\text{BH}\)) and found that their masses are \(M_\text{BH}\) \( < 10^{8.5}\,M_\odot\), that according to\cite{weinberger2018supermassive}, these black holes are most effective in the thermal mode that is the dominant mass growth channel for our void galaxies' SMBH (\cite{zinger2020ejective})(See Figure \ref{fig:stellar_mass_2comparison}).

We also analyzed the evolutionary history of the main parameters of these void galaxies from redshift \(Z = 0\) to \(Z = 2\) ( \(last \approx 12.5Gyr\)) to explore any discrepancy and any changes in redshifts. We found that quenched void galaxies have higher SFR in high redshift than star-forming galaxies and decreased SFR,  in lower redshifts. These patterns are also seen in all stellar mass ranges, but star-forming void galaxies have star formation rates uniformly over time (See Figure \ref{fig:stellar_mass_3comparison}).
We also explored the stellar mass assembly of our void galaxies, and we showed that quenched void galaxies form 50\% of their mass in all stellar mass bins(more pronounced in lower stellar mass) earlier than star-forming void galaxies (\( \approx 11Gyr\) ago)(See Figure \ref{fig:stellar_mass_5comparison}). We could conclude that void-quenched galaxies likely formed earlier in cosmic history, rapidly consuming their available gas during an initial phase of heightened star formation. This early activity led to the depletion of their gas reservoirs, resulting in a gradual decline in star formation rates over time. These galaxies exhibited significantly reduced star formation by low redshift and were effectively quenched by \( z = 0 \) (\cite{behroozi2019universemachine}). These results were also observationally studied by \cite{dominguez2023galaxies} that classified the star-formation histories (SFHs) of void galaxies into two types: the short-timescale SFH
(ST-SFH) is characterized by high star formation at the earliest time and decreased star formation later in their lives, while the long-timescale SFH (LT-SFH) has star formation happening more uniformly over time. Our study also examined the histories of major mergers in void galaxies across different time scales of their evolution. In Figure 12, the heatmap reveals that at late time, star-forming void galaxies exhibit a higher fraction of systems with at least one major merger compared to their quenched counterparts.  These findings suggest that ongoing mergers are critical in sustaining star formation and delaying quenching in star-forming galaxies. In contrast, quenched galaxies, having experienced their peak merger activity at earlier epochs, show minimal recent merger contributions, reflecting their transition into more passive systems. These results can be the reason for the higher star formation rate of quenched galaxies in earlier times. 

\begin{figure}[t] 
    \centering
    \includegraphics[width=1\columnwidth, keepaspectratio]{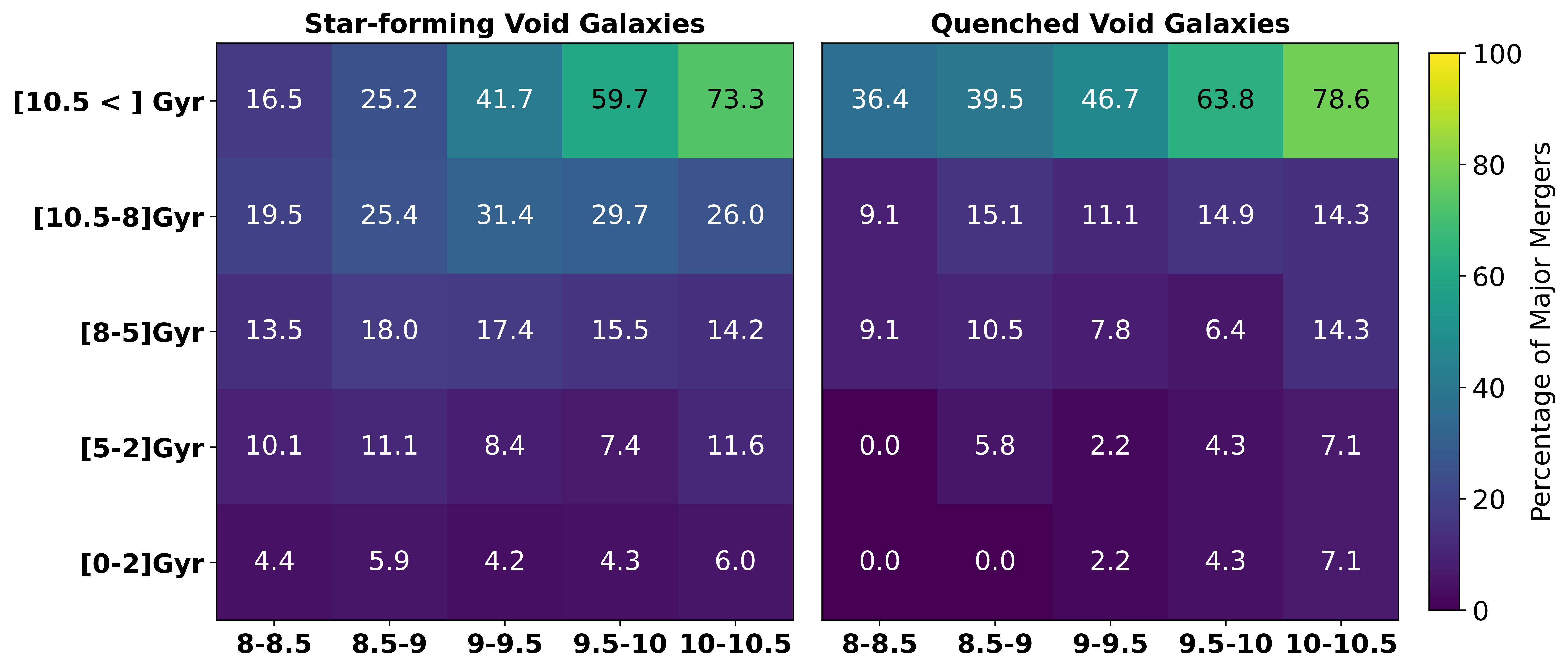} 
  \caption{
Percentage of major mergers for star-forming and quenched void galaxies across time scales: present to 2 Gyr, 5–2 Gyr, 8–5 Gyr, 10.5–8 Gyr, and before 10.5 Gyr. Stellar mass bins (\(\log_{10} M_{\odot}\)) are on the x-axis, and colors indicate the percentage of galaxies with at least one major merger per bin.
}

    \label{fig:stellar_mass_comparison}
\end{figure}
We analyzed the void galaxies in more detail and found that quenched void galaxies reside in higher dark matter halos over cosmic time and in all stellar mass ranges(See Figure \ref{fig:stellar_mass_4comparison}). More dark matter by quenched void galaxies may be efficient in accelerating 
 star formation efficiency (the fraction of gas converted into stars on
a free-fall time) and lead to rapid gas consumption, that are more effective in higher redshifts(\cite{boylan2024accelerated}). 

We also studied supermassive black holes' accretion rates (\(\dot{M}_\text{BH}\))  across redshift and found their behavior is the same as SFR for our quenched and star-forming void galaxies (See Figure \ref{fig:stellar_mass_4comparison}). While the thermal feedback mode of SMBHs does not effectively reduce star formation rate(\cite{weinberger2018supermassive}) \(\dot{M}_\text{BH}\) - SFR relation indicates that the gas present in the galaxy “prefers” to fall into the supermassive black hole of void galaxies and effects on star formation rate and accretion rate of black holes simultaneously.  In quenched void galaxies, reduced star formation rates (SFR) and black hole accretion rates (\(\dot{M}_\text{BH}\)) at low redshifts can be attributed to their earlier formation consuming their gas reservoirs and the subsequent lack of adequate gas inflows. Interactions and mergers between galaxies are known to trigger large-scale nuclear gas inflows, which is a prerequisite for the growth of central black holes by accretion(\cite{hernquist1989tidal}; \cite{barnes1992dynamics} ; \cite{capelo2015growth} ). A positive correlation between the star formation rate (SFR) and the supermassive black hole (SMBH) accretion rate was studied by \cite{byrne2023interacting}. The study concluded that gas-rich major mergers exhibit the strongest correlation between SFR and SMBH accretion rate (\(\dot{M}_{\mathrm{BH}}\)), with the majority showing enhancements in both.

In our study, we examined the ex-situ stellar mass fraction (\( f_{\text{acc}} \)), 
which measures the average amount of stellar mass that a
galaxy accretes per unit of time through mergers with other galaxies. As discussed in \cite{rodriguez2017role} at any fixed stellar mass –\( f_{\text{acc}} \) is negatively correlated with gas-rich mergers. Our findings revealed that ex-situ stellar mass fractions typically stay under 0.1 for most void galaxies (See Figure \ref{fig:UVJ_diagram}). This suggests that in-situ star formation is the primary driver of their growth, and wet mergers dominate in void environments and the vast majority of the merger-acquired
baryons are gaseous. Our results in ex-situ stellar mass fraction (\( f_{\text{acc}} \)) and the gas fraction values of void galaxies across redshifts (See Figure \ref{fig:stellar_mass_5comparison}) indicate mergers in star-forming and quenched void galaxies are gas-rich. Mergers in low-dense environments are also studied in some papers; an observational study by \cite{sureshkumar2024galaxy} showed that galaxy mergers prefer under-dense environments of the cosmic web in the local Universe, or \cite{lin2010wet} showed that wet mergers mainly occur in low-dense environments and dry mergers occur in denser environments. Also, \cite{ceccarelli2024galaxy}  found that the efficiency of star formation associated with galaxy interactions is significantly larger in cosmic void pairs. 

This paper investigated the statistical merger rates of star-forming and quenched void galaxies across cosmic time. We found that in void galaxies, the fraction of mergers decreases significantly from high redshift to low redshift, reflecting the isolated environments of voids where galaxies have fewer neighbors and interactions become less frequent over time. Mergers—both major and minor—were more common in the early universe when galaxies were denser and more dynamically active, but their rates declined as void regions became more isolated due to cosmic expansion. Star-forming galaxies exhibit a higher fraction of mergers (major, minor, and total) than quenched galaxies. Major mergers in quenched galaxies show a more pronounced decline at earlier times than present, while the difference in minor merger fractions between star-forming and quenched galaxies becomes relatively small at lower redshifts. (See Figure \ref{fig:stellar_mass_6comparison} and  \ref{fig:stellar_mass_7comparison}). 
Based on our results, star-forming void galaxies, on average, experienced their last merger more recently than quenched void galaxies. This trend is statistically consistent across all mass ranges and for all types of mergers, highlighting a significant difference in the merger histories of star-forming and quenched void galaxies. Additionally, star-forming has a higher fraction of one merger in all mass ratios of mergers in the last 5 Gyr and a higher rate of multiple mergers in the last 8 Gyr statistically in all stellar mass ranges. These results show that mergers have a crucial role in different histories of star formation rates of star-forming and quenched void galaxies, especially in the last 8 Gyr, where they have had strong different trends.  (See Figure \ref{fig:stellar_mass_10comparison} and  \ref{fig:stellar_mass_11comparison}).

Star formation rate enhancements of galaxies after mergers studied by \cite{hani2020interacting} and \cite{byrne2023interacting}. They demonstrated that the enhancements persisted for up to $\sim$500 Myrs
 after the merger independent of the merger mass ratio, consistent with previous simulation studies (\cite{di2008frequency}) and observational estimates (\cite{wild2010timing} ). Also, in gas-rich environments like voids and for stellar masses below \(10^{10.5} \, M_\odot\), this effect could be more pronounced. Therefore, mergers can sustain star formation in star-forming galaxies at low redshifts due to recent mergers and multiple mergers occurring at these times.

An analysis of non-mergers and systems with at least one merger among star-forming and quenched void galaxies reveals distinct patterns in the impact of mergers. In low-mass quenched galaxies (\(10^8 - 10^9 \, M_\odot\)), mergers exert minimal influence, with similar distributions of star formation rate (SFR), specific star formation rate (sSFR), star formation efficiency (SFE), and gas fraction observed in both merger and non-merger populations. For high-mass quenched galaxies (\(10^9 - 10^{10.5} \, M_\odot\)), mergers result in slight enhancements in SFR and SFE, although differences in gas fractions remain statistically insignificant. In stark contrast, mergers have a pronounced effect on star-forming galaxies, significantly boosting SFR, SFE, and gas retention in low- and high-mass systems. These findings underscore the critical role of mergers in sustaining and enhancing star formation in void galaxies, particularly in star-forming systems, where mergers drive prolonged dynamical activity and gas accretion. (See Figure \ref{fig:01stellar_mass_comparison} and Table \ref{tab:ks_minor_major_comparison_v3}.)

In summary, in our analysis, quenched void galaxies may consume their gas rapidly in earlier epochs due to their earlier formation and residence in higher-mass dark matter halos, leading to their transition into passive systems in low redshifts. In contrast, star-forming galaxies formed later, and owing to statistically higher merger rates at lower redshifts in all types of mergers across various stellar mass ratios, they maintain a relatively uniform star formation rate across redshifts. This sustained star formation is further enhanced by the influx of gas and dynamical activity triggered by recent mergers. Gas-rich environments in void galaxies also provide favorable conditions for prolonged star formation, as these mergers replenish the gas supply and stimulate star formation even at low redshifts. Therefore, we can conclude mergers significantly influence the star formation rate of void galaxies, shaping their evolutionary pathways and sustaining activity in star-forming systems.

\bibliographystyle{aa}  
\bibliography{aanda}  

\end{document}